\documentclass[prd,aps,nofootinbib,showpacs,floats,letterpaper,floatfix,groupedaddress,eqsecnum]{revtex4}

\usepackage{amssymb,amsmath}
\usepackage[dvips]{graphicx}

\usepackage{dcolumn,epsfig}

\def\nn{\nonumber}

\def\lisa{{\em LISA}}

\def\be{\begin{equation}}
\def\ee{\end{equation}}
\def\beq{\begin{eqnarray}}
\def\eeq{\end{eqnarray}}

\def\IL{\relax{\rm I\kern-.18em L}}

\def\nn{\nonumber}
\def\f{\frac}

\newlength{\sizeonefig}
\newlength{\sizetwofig}
\begin{document}

\title{Quasinormal ringing of Kerr black holes: \\
The excitation factors}

\author{Emanuele Berti} \email{berti@wugrav.wustl.edu}
\affiliation{McDonnell Center for the Space Sciences, Department of
Physics, Washington University, St.  Louis, Missouri 63130, USA}

\author{Vitor Cardoso} \email{vcardoso@phy.olemiss.edu}
\affiliation{Dept. of Physics and Astronomy, The University of
Mississippi, University, MS 38677-1848, USA \footnote{Also at Centro
de F\'{\i}sica Computacional, Universidade de Coimbra, P-3004-516
Coimbra, Portugal}}

\date{\today}

\begin{abstract}

  Distorted black holes radiate gravitational waves. In the so-called ringdown
  phase radiation is emitted as a discrete set of complex quasinormal
  frequencies, whose values depend only on the black hole's mass and angular
  momentum. Ringdown radiation could be detectable with large signal-to-noise
  ratio by the Laser Interferometer Space Antenna (\lisa). If more than one
  mode is detected, tests of the black hole nature of the source become
  possible. The detectability of different modes depends on their relative
  excitation, which in turn depends on the cause of the perturbation (i.e. on
  the initial data). A ``universal'', initial data-independent measure of the
  relative mode excitation is encoded in the poles of the Green's function
  that propagates small perturbations of the geometry (``excitation
  factors'').  We compute the excitation factors for general-spin
  perturbations of Kerr black holes. We find that for corotating modes with
  $l=m$ the excitation factors tend to zero in the extremal limit, and that
  the contribution of the overtones should be more significant when the black
  hole is fast rotating. We also present the first analytical calculation of
  the large-damping asymptotics of the excitation factors for static black
  holes, including the Schwarzschild and Reissner-Nordstr\"om metrics. This is
  an important step to determine the convergence properties of the quasinormal
  mode expansion.
\end{abstract}


\maketitle

\section{Introduction}

Distorted black holes emit gravitational radiation as a discrete sum of
quasinormal modes (QNMs), damped oscillations whose frequencies and damping
times depend only on the black hole's mass and angular momentum. The QNM
frequencies scale with the inverse of the black hole's mass. Therefore, the
optimal sensitivity bandwidth of each detector determines the mass range that
can be detected.  The dominant QNM frequency for quadrupolar radiation from a
Schwarzschild black hole is $f=1.207\times 10^{-2} (10^6 M_\odot/M)$~Hz. As a
consequence, the collapse of Population-III stars with mass $M\gtrsim 260
M_\odot$ or $25 M_\odot \lesssim M\lesssim 40 M_\odot$ forms massive black
hole remnants that could be detectable by ground-based, high-frequency
gravitational wave interferometers such as the advanced Laser Interferometer
Gravitational-Wave Observatory (LIGO) \cite{advLIGO} and Virgo \cite{Virgo}.
The space-based Laser Interferometer Space Antenna (\lisa), being sensitive in
the frequency band $\sim 10^{-5}-10^{-1}$~Hz, will detect the gravitational
radiation emitted by oscillating black holes of mass $10^5 M_\odot\lesssim
M\lesssim 10^8-10^9~M_\odot$ with large signal-to-noise ratio (SNR) throughout
the observable Universe \cite{FH,BCW}. With large SNR come precise
measurements of the source parameters, and an extraordinary opportunity to
study the physics of massive black holes \cite{gwdaw10}.

Since the \lisa~SNR can be very large, more than one mode could be detected in
the ringdown waveform. Such a detection would allow an unprecedented test of
the black hole nature of the source. The basic idea is quite simple.  In
general relativity the complex QNM frequencies form a discrete set
$\omega_{lmn}$ classified by three integers: the indices $(l,m)$ come from the
separation of the angular dependence of the perturbations, and the index $n$
labels frequencies by the magnitude of their imaginary part (large $n$ means
large imaginary part and short damping time).  Because of the ``no-hair
theorem'', QNM frequencies depend only on the mass and angular momentum of a
Kerr black hole. Roughly speaking, the measurement of one complex frequency
(two observables) provides us with a determination of the mass and angular
momentum of the black hole, and the measurement of the second frequency allows
a consistency check with the general relativistic predictions
\cite{BCW,gwdaw10,dreyer,bcboson}.

The detectability of different QNMs depends on the relative QNM excitation,
which in turn is determined by the cause of the perturbation (i.e. by the
initial data). Given a detection, the resolvability of different modes depends
on the nature of the multimode ringdown waveform. Two scenarios are possible
\cite{BCW}. If $l\neq l'$ or $m\neq m'$ (the modes have different angular
dependence) the angular scalar product between the modes is zero to a good
approximation \cite{BCC}, and the SNR can be expressed as a sum in quadrature
of the single-mode SNRs. In \cite{BCW} we called these multimode ringdown
waveforms {\it quasi-orthonormal}. If instead $l=l'$ and $m=m'$, but $n\neq
n'$ (we look at different overtones with the same angular dependence) the
angular scalar product is very close to unity, and mixed terms appear in the
SNR calculation and in calculations of parameter estimation accuracy. We may
call these waveforms {\it quasi-parallel}.

A quantitative estimate of the relative QNM excitation for quasi-orthonormal
waveforms will require numerical relativistic simulations and realistic
initial data for the merger \cite{BCW2}. However, for quasi-parallel waveforms
perturbation theory (combined with a good approximation to the initial data
for ringdown) can provide useful information on the relative excitation of
different overtones.  In this paper, using perturbation theory, we develop
tools to estimate the relative excitation of different overtones of Kerr black
holes for general classes of initial data. We also try to gain some
theoretical insight into the initial-data dependence of the results for
realistic mergers, considering simple classes of initial data (such as
localized spikes and gaussians) as model problems.

Our work is the first application to gravitational perturbations of Kerr black
holes of a theoretical framework developed over the years by different
authors. Following the pioneering numerical analysis by Vishveshwara
\cite{vish}, who studied the scattering of gaussian wave pulses on the
Schwarzschild background, a number of studies investigated the analytical
structure of the Green's function that propagates small perturbations in black
hole geometries. Our own analysis is based on the formalism developed by
Leaver in the eighties \cite{lePRS,leJMP,lePRD}, but the Green's function in
the Schwarzschild background has been studied by many authors
\cite{sunprice88,nollertschmidt,nollertprice,nils95,nils97}.  More recently
Glampedakis and Andersson extended the analysis to {\it scalar} perturbations
of Kerr black holes \cite{kostasnils,quickdirty}. Here we carry out the first
study of general-spin perturbations of Kerr black holes, including the
physically most important case of gravitational perturbations.


The main result emerging from
\cite{lePRS,leJMP,lePRD,sunprice88,nollertschmidt,nollertprice,
  nils95,nils97,kostasnils,quickdirty} is that a ``universal'' (initial-data
independent) measure of the relative QNM excitation is encoded in the poles of
the Green's function that propagates small perturbations of the geometry.
These ``universal'' {\it quasinormal excitation factors} (QNEFs) are defined
in Eq.~(\ref{excfactors}) below. They depend only on the Kerr geometry, not on
the details of the perturbation.  When combined with a knowledge of the
initial data they can be used to compute the so-called {\it quasinormal
  excitation coefficients} defined in Eq.~(\ref{exccoeffs}), which are a
concrete measure of the QNM content of a waveform.

The paper is organized as follows.
In Sec.~\ref{string}, to develop some physical intuition, we consider a very
simple physical system: a vibrating string with fixed ends. This part provides
a useful pedagogical introduction to the QNM excitation problem, and may be
skipped by readers familiar with the topic. We consider the Green's function
solution of the vibrating string equation for generic initial data, identify
the normal modes of the system as poles of the Green's function and show the
importance of initial data to determine the excitation of the modes.
In Sec.~\ref{bholes} we show that many features of the vibrating string
problem carry over to black hole perturbation theory, stressing the main
differences between normal mode expansions and QNM expansions. We also
anticipate some results on the convergence of the QNEFs, which are presented
in more detail in Appendix \ref{app:asymptBn}.
In Sec.~\ref{Bn} we outline our calculation of the QNEFs, present the
numerical results and compute the response of a Kerr black hole to localized
and gaussian initial data. Technical details, as well as a discussion of the
numerical accuracy of our calculations, are relegated to Appendix
\ref{app:tech}.  Appendix \ref{app:SN} clarifies the relation between the
Teukolsky and Sasaki-Nakamura (SN) formulations of the Kerr perturbation
equations, and between the corresponding QNEFs.

\section{A pedagogical example: a vibrating string}
\label{string}

Some key features of the black hole perturbation problem we address in this
paper are exemplified by a very simple system: a vibrating string with fixed
ends.  To simplify the mathematics we pick units so that the velocity of the
waves in the string $c=1$, and consider a string of length $\pi$.  Then any
disturbance of the string obeys the wave equation
\be
\frac{\partial^2u}{\partial t^2}=\frac{\partial^2u}{\partial x^2}\quad
{\rm on}\quad 0\leq x\leq \pi\,,
\label{flatwave}
\ee
with $u(t,0)=u(t,\pi)=0$. The general solution of this problem is easily
verified to be
\be u(t,x)=\sum_{n=1}^{\infty}
\left (\bar C_n \cos{nt}+\bar C_n' \sin{nt}\right )\sin{(n x)}\,,
\label{gensolstring} \ee
where we used an overbar to avoid confusion with the {\it quasinormal}
excitation coefficients $C_n$, as defined in Eq.~(\ref{exccoeffs}) below. In
Fourier language we say the general solution is a superposition of normal
modes with sinusoidal dependence on $x$ and $t$, labeled by an integer $n$.
Each mode has frequency $\omega=n$.

For the general solution to be useful we must determine the constants $\bar
C_n$ and $\bar C_n'$, that is, we must determine the {\it contribution of each
  individual mode}. This contribution can easily be computed once we are given
{\it initial data}, namely the initial configuration $u(0,x)\equiv u_0(x)$ and
velocity profile $\partial_t u(0,x) \equiv v_0(x)$ of the string. Indeed,
consider (\ref{gensolstring}) and its first derivative, both evaluated at
$t=0$. Multiplying both sides by $\sin nx$ and integrating on $(0\,,\pi)$
%
%
we get
\beq
\bar C_n=\frac{2}{\pi}\int_0^{\pi}u_0(x)\,\sin nx \,dx\,, \qquad
\bar C_n'=\frac{2}{n\pi}\int_0^{\pi}v_0(x)\,\sin nx \,dx\,, \label{initFS}
\eeq
which completely specifies the solution.

In more general situations it is not possible to find closed-form elementary
solutions satisfying some given boundary conditions (in our vibrating string
example, the fixed ends condition).  However, an elegant formal solution can
be obtained using Green's functions. Let us consider a slight generalization
of Eq.~(\ref{flatwave}):
\be
\frac{\partial^2 u}{\partial x^2}- \frac{\partial^2 u}{\partial t^2}
-V(x)\,u={\cal S}\,,\label{waveeq-gen}
\ee
where we introduced a potential $V(x)$ and a source ${\cal S}$ representing,
say, external forces acting on the system (for a free vibrating string
$V(x)={\cal S}=0$).  We define the Laplace transform of $u(t,x)$
as\footnote{The usual Laplace variable $s=-i\omega$. We prefer to use $\omega$
  for notational consistency with previous work by Leaver \cite{lePRD} and
  Andersson \cite{nils95}. Our transform is well defined as long as ${\rm
    Im}(\omega)\geq c$.}
\be
{\cal L}u(t,x)
\equiv {\hat u}(\omega\,,x)=
\int_{t_0}^{\infty} u(t,x)e^{i\omega t} dt\,,
\ee
In terms of the Laplace transform, the original field can be written as
\be
u(t,x)=\frac{1}{2\pi}\int_{-\infty +ic}^{\infty+ic} {\hat u}(\omega\,, x)e^{-i\omega t} d\omega\,.\label{originalu}
\ee
Using the elementary property ${\cal L}\left[\f{\partial u(t,x)}
{\partial t}\right]=-i\omega {\cal L}u-e^{i\omega t_0}u(x,t_0)$, the Laplace
transformation of (\ref{waveeq-gen}) leads to
\be
\frac{\partial^2{\hat u}}{\partial x^2}+\left [\omega^2-V(x)\right ]{\hat u}=I(\omega\,,x)\,,\label{waveeq}
\ee
where
\be\label{Iw}
I(\omega\,,x)=e^{i\omega t_0} \left [i\omega u(t,x)-\frac{\partial u(t,x)}{\partial t} \right ]_{t=t_0}+\hat{\cal S}\,.
\ee
This equation is formally solved with the use of a Green's function $G(x,x')$
such that
\be
\frac{\partial^2{\hat u}}{\partial x^2}+\left [\omega^2-V(x)\right ]G(x\,,x')=\delta(x-x')\,.\label{greenhomo}
\ee
In terms of the Green's function the solution is simply given by
\be
{\hat u}=\int I(\omega\,,x')G(x\,,x')dx'\,. \label{solgreen}
\ee
Suppose we know the Green's function. Then the previous equation shows that,
given $I(\omega\,,x')$ (which means, in the absence of external forces, given
initial data) we can determine, at least in principle, the solution.

\begin{figure*}[ht]
\begin{center}
\begin{tabular}{cc}
\epsfig{file=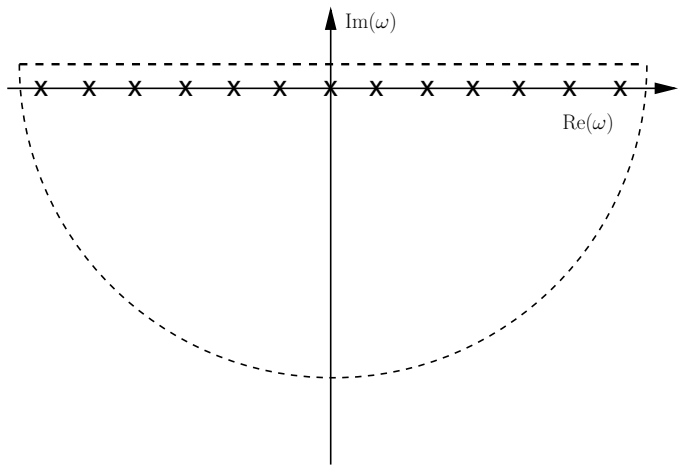,width=9cm,angle=0}
\end{tabular}
\end{center}
\caption{Integration contour for the vibrating string problem. Crosses mark
  zeros of the Wronskian $W$, corresponding to the normal frequencies of the
  system.}
\label{fig:contour1}
\end{figure*}

There is a general prescription to construct the Green's function
\cite{morsefeshbach}. Find two linearly independent solutions of the
homogeneous equation, say ${\hat u}_1(\omega\,,x)$ and ${\hat
  u}_2(\omega\,,x)$, each satisfying one of the required boundary conditions:
for the vibrating string these solutions would be such that ${\hat
  u}_1(\omega\,,0)=0$, ${\hat u}_2(\omega\,,\pi)=0$. The Green's function is
then
\be
G(x\,,x')=\frac{1}{W}\left\{ \begin{array}{ll}
             {\hat u}_1(x){\hat u}_2(x')   & \mbox{if $x\leq x'$}\,,\\
             \\
              {\hat u}_1(x'){\hat u}_2(x)    & \mbox{if $x'\leq x$}\,,
\end{array}\right.
\label{g1}
\ee
where we defined the Wronskian between the two solutions
\be
W=\frac{\partial {\hat u}_1}{\partial x}{\hat u}_2-
{\hat u}_1\frac{\partial {\hat u}_2}{\partial x}\,,
\ee
which for equations of the type (\ref{waveeq}) is a constant.  For the
vibrating string the homogeneous solutions are elementary functions: ${\hat
  u}_1=\sin \omega x$, ${\hat u}_2=\sin\omega (x-\pi)$, and the Green's
function
\be
G(x\,,x')=\left\{ \begin{array}{ll}
            -\frac{\sin \omega x \sin \omega (x'-\pi)}{\omega \sin \omega \pi}    & \mbox{if $x\leq x'$}\,,\\
             \\
             -\frac{\sin \omega x' \sin \omega (x-\pi)}{\omega \sin \omega \pi} & \mbox{if $x'\leq x$}\,.
\end{array}\right.
\label{g2}
\ee
Notice that the Wronskian $W=-\omega \sin \omega \pi$ is zero at $\omega = n$
with $n$ integer, that is, at the {\it normal frequencies} of the system. Near
the poles $\omega=n$ we have $W \simeq -[n \pi \cos(n\pi)] \,(\omega-n)$.
Corresponding to zeros of the Wronskian (which are also {\it poles of the
  Green's function}: see below) the two solutions ${\hat u}_1$ and ${\hat
  u}_2$ are no longer independent: they satisfy both boundary conditions
simultaneously. In fact, setting $\omega=n$ in (\ref{g2}) we can see that
${\hat u}_1$ and ${\hat u}_2$ coincide, and correspond to the normal modes of
the system. Using (\ref{originalu}), (\ref{Iw}) and (\ref{solgreen}) and
setting for simplicity $t_0={\cal S}=0$ we get
\be u(t\,,x)=\frac{1}{2\pi}\int dx' d\omega \left [i\omega u_0(x')-v_0(x')\right ]G(x\,,x')e^{-i\omega
t}\,.\label{xxx}
\ee
The $\omega-$integral can be performed by closing the contour of integration.
We choose the contour depicted in Fig.~\ref{fig:contour1}.

If ${\hat u}_1(x)\,,{\hat u}_2(x)$ are analytic and have no essential
singularities inside the contour\footnote{Both conditions are met in the case
  of a vibrating string, but both are {\it violated} when we deal with black
  hole QNMs. For black holes, the Green's function essential singularity at
  the origin gives rise to tails, and the integral over the half-circle at
  infinity is responsible for the early-time response of the black hole. See
  discussion after Eq. (\ref{zz}).}, the poles of the Green's function are all
due to zeros of the Wronskian. In this way we get for the integral in
(\ref{xxx}),
\beq
u(t\,,x)
&=&\frac{1}{2\pi}\int dx' d\omega \left [i\omega u_0(x')-v_0(x')\right ]
G(x\,,x')e^{-i\omega t} \nn \\
&=&-i \sum_n \f{1}{n\pi} \left\{ \int dx' \sin nx' \left [in
    u_0(x')-v_0(x')\right]\right\} e^{-in t} \sin nx\nn\\
&&+\frac{1}{2\pi} \int_{HC} d\omega
\int dx'\left [i\omega u_0(x')-v_0(x')\right ] G(x\,,x')e^{-i\omega t}
\,, \label{closecontour}
\eeq
where ``HC'' in the second term means that the integration should be performed
along the half-circle. Not surprisingly, taking the real part of the right
hand side we recover the result (\ref{gensolstring}) with expansion
coefficients given by (\ref{initFS}).

The bottom line of this discussion is that the general solution depends
crucially on two elements: (i) the residues of the Green's function evaluated
at the poles (that is, at the normal frequencies); (ii) the function
$I(\omega\,,x)$, which (in the absence of initial forces) is nothing but the
initial data. By inspection, the net result for the field can be expressed as
a sum of the form
\be
u(t\,,x)\propto \sum_{n} \frac{J_n(t\,,x)}{\partial_{\omega} W}
u_n(\omega_n\,,x) e^{-i\omega_n t}\,,
\ee
where $\omega_n=n$ is a normal frequency, $J_n(t\,,x)=\int dx'
I(\omega_n\,,x')\,u_n(\omega_n\,,x')$ and $u_n(\omega_n\,,x)=\sin nx$ is a
normal mode wavefunction (that is, any of the homogeneous solutions evaluated
at the normal frequency $\omega_n=n$). We will see below that a similar result
holds for the ringdown of Kerr black holes.

\subsection{Effect of initial data}

In the previous section we pointed out that initial data play a crucial role
to determine the excitation of the normal modes of a system. For illustration,
below we consider three simple examples that will be useful in the following
to understand, by analogy, the initial-data dependence of the excitation of a
Kerr black hole.

Suppose first that we have an (initially stationary) plucked string: the
string's initial profile is a triangle of height $h$ with a vertex at $x^S$,
i.e. $v_0(x)=0$ and
\be
u_0(x)=\left\{ \begin{array}{ll}
             \frac{x h}{x^S}   & \mbox{if $0\leq x\leq x^S$}\,,\\
             \\
             \frac{h(\pi-x)}{\pi-x^S}  & \mbox{if $x^S\leq x\leq\pi$}\,.
\end{array}\right.
\label{Heaviside}
\ee
Stationarity of the initial data implies that $\bar C_n'=0$, so that
\be u(t,x)=\sum_{n=1}^{\infty}\bar C_n \cos nt \sin nx\,, \quad
\bar C_n=\frac{2h}{n^2}\frac{\sin n x^S}{x^S(\pi-x^S)}
\ee
Notice that modes having a node at $x^S$, where the string is plucked, are not
excited ($\bar C_n=0$). Notice also that the "excitation factors" $\bar C_n$
decrease as $1/n^2$.

As a second example take stationary, localized ``$\delta$-function'' initial
data of the form
\be u_0(x)=\delta(x-x^S)\,,\quad v_0(x)=0\,.\ee
The excitation factors are trivially computed:
\be \bar C_n=\frac{2}{\pi}\sin{(n x^S)}\,. \label{pointstring}\ee
It is apparent that all modes are excited to a comparable amplitude except for
modes with a node at the plucking point, which are not excited at all.

Our third and last example are stationary, gaussian initial data:
\be u_0(x)=e^{-b(x-x^S)^2} \,,\quad v_0(x)=0\,.\ee
For large $b$ the gaussian is strongly peaked at $x=x^S$, in which case the
contribution to the integral outside of $(0,\pi)$ can be ignored and we have
\be
\bar C_n=\frac{2}{\pi}\int_0^{\pi}u_0(x)\,\sin nx \,dx
\simeq \frac{2}{\pi}\int_{-\infty}^{\infty}e^{-b(x-x^S)^2}\sin nx \,dx
=\frac{2\sin (nx^S)}{\sqrt{\pi b}}e^{-n^2/(4b)}
\label{stringexc}
\ee
Therefore a mode with given $n$ is maximally excited when the width of the
gaussian satisfies the condition $b=n^2/2$. A similar result will be seen to
hold for gaussians exciting Kerr black holes.

The basic lesson we learn from these examples is that the excitation of a
system is very sensitive to the initial data. More specifically, whether a
given mode is excited or not depends strongly on the point where we excite the
system (``pluck the string'').

\section{Oscillating black holes}
\label{bholes}

In the SN formalism \cite{sasakinakamura,SN2,hughes,tagoshi}, perturbations of
a Kerr black hole induced by a spin-$s$ field are described by a single
function $X^{(s)}(t,r)$ whose Laplace transform satisfies
\be
\frac{d^2{\hat X^{(s)}}(\omega\,,r)}{dr_*^2}
+V_{SN}{\hat X^{(s)}}(\omega\,,r)=I(\omega\,,r)\,,
\label{potbh1}
\ee
where the effective potential $V_{SN}$ depends both on the radial coordinate
and on the frequency. The function $I(\omega\,,r)$ is a linear combination of
the SN function $X^{(s)}(t_0\,,r)$ and its time derivative $\dot
X^{(s)}(t_0\,,r)$ at time $t_0$ (see below for the explicit expression for
scalar perturbations).  The tortoise coordinate $r_*$ is defined by the
condition
\be
\f{dr_*}{dr}=\f{r^2+a^2}{\Delta}\,,
\ee
ranging from $-\infty$ (the location of the event horizon) to $+\infty$
(spatial infinity). We use Boyer-Lindquist coordinates and follow Leaver's
choice of units, setting $G=c=2M=1$. In Leaver's units the angular momentum
per unit mass $a$ is such that $0\leq a \leq M=1/2$, and the horizon function
$\Delta= r^2-r+a^2$. Sometimes we will present our results in terms of the
more familiar dimensionless angular momentum $j=2a$, such that $0\leq j \leq
1$. The class of problems that fit in this description include any massless
field in the Kerr geometry, including gravitational, electromagnetic and
scalar fields \cite{commentsn}.

The main difference with the vibrating string example is that our system is
not conservative: waves can escape to infinity. For this reason an expansion
in normal modes is not possible (see
\cite{nollertschmidt,kokkotasnollertreview,lePRD} for extensive discussions of
this point). Wave propagation is also complicated by backscattering off the
background curvature, which is responsible for tail effects \cite{price}.
Despite these complications it can be shown that the poles of the Green's
function (now located at complex frequencies corresponding to the QNMs) still
play an important role in the evolution.

\begin{figure*}[ht]
\begin{center}
\begin{tabular}{cc}
\epsfig{file=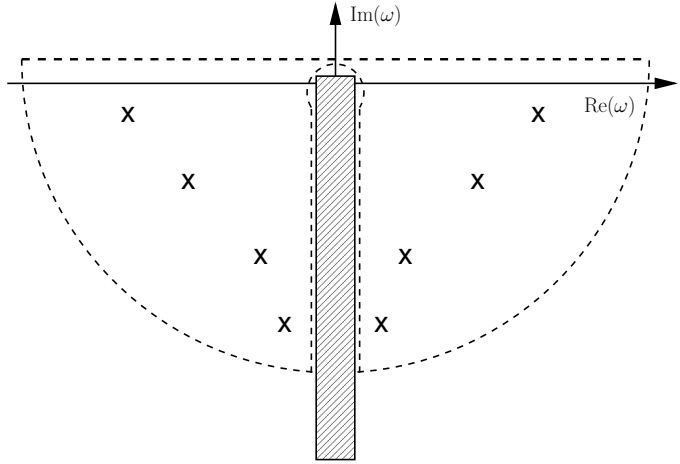,width=9cm,angle=0}
\end{tabular}
\end{center}
\caption{Integration contour to invert Eq.~(\ref{zz}). The shaded area is the
  branch cut and crosses mark zeros of the Wronskian $W$ (the QNM
  frequencies).}
\label{fig:contour2}
\end{figure*}

The QNM contribution can be isolated from other features of the signal, such
as the late-time tail, using the Green's function technique
\cite{lePRD,nils95}. First one defines a solution of the homogeneous equation
having the correct behavior at the horizon (only in-going waves),
\beq
& & \lim_{r \to r_+} {\hat X^{(s)}}_{r_+}\sim e^{-i(\omega-m\Omega) r_*}\,,
\label{asrplus}\\
& & \lim_{r \to \infty }{\hat X^{(s)}}_{r_+}\sim A_{\rm in}(\omega)e^{-i\omega r_*}+A_{\rm out}(\omega)e^{i\omega
r_*}\,, \eeq
where $r_\pm=[1\pm (1-4a^2)^{1/2}]/2$, and a second solution ${\hat
  X^{(s)}}_{\infty_+}$ behaving as $e^{i\omega r_*}$ for large values of $r$.
Since the Wronskian $W=2i\omega A_{\rm in}$ we can express the general
solution as \cite{nils95}
\be {\hat X^{(s)}}(\omega\,,r)={\hat X^{(s)}}_{\infty_+}\int_{-\infty}^{r_*}\frac{I(\omega\,,r){\hat X^{(s)}}_{r_+}}{2i\omega A_{\rm
in}}dr_*' +{\hat X^{(s)}}_{r_+}\int_{r_*}^{\infty}\frac{I(\omega\,,r){\hat X^{(s)}}_{\infty_+}}{2i\omega A_{\rm in}}dr_*' \,.
\ee
To proceed we make the astrophysically reasonable assumption that the observer
is located far away from the black hole. If the initial data have compact
support and this support is entirely located closer to the black hole with
respect to the observer (this is basically a ``no-incoming radiation from
infinity'' condition), a good approximation will be
\be {\hat X^{(s)}}(\omega\,,r)\simeq \frac{e^{i\omega r_*}}{2i\omega A_{\rm in}}\int_{-\infty}^{\infty}
I(\omega\,,r){\hat X^{(s)}}_{r_+}dr_*' \,.\label{zz}
\ee
As explained in \cite{lePRD,nils95,nollertthesis}, when we invert this
expression to get the solution in the time domain we get (as in the vibrating
string case) a contribution from the poles of the Green's function.  This
contribution can again be isolated by closing the path of integration, as was
done in Eq.~(\ref{closecontour}). An important difference that distinguishes
black hole spacetimes is that there is now an essential singularity at
$\omega=0$ and a branch cut extending from the singularity to $-i \infty$
\cite{nollertthesis,jensen,ching}. To prevent the essential singularity from
lying inside the integration contour we must modify slightly the contour of
Fig.~\ref{fig:contour1}. We place a branch cut along the negative
imaginary-$\omega$ axis and split the half circle at $|\omega|\rightarrow
\infty$ into two quarter circles. The new contour is shown in
Fig.~\ref{fig:contour2}.
The poles in the complex frequency plane are the zeros of $A_{\rm in}$: they
correspond to perturbations satisfying {\it both} in-going wave conditions at
the horizon and out-going wave conditions at infinity, that is (by definition)
to QNMs. QNM frequencies have negative imaginary part. Since we assume a
Fourier dependence of the form $X^{(s)}(t,r)\sim e^{-i \omega t}$ this means
that there are no exponentially growing modes.
There is an infinity of QNMs, which are usually sorted by the magnitude of
their imaginary part and labelled by an integer $n$.

Extracting the QNM contribution\footnote{There are also other contributions to
  the signal. The essential singularity at the origin ($\omega = 0$) gives
  rise to the tail of the time evolved wavefunction; the integral over the
  quarter circles at infinite frequency produces the early time response of
  the black hole. Since our main focus is the QNM contribution we discard
  these two terms in the integral. For more details we refer the reader to
  \cite{lePRD,nollertthesis,ching}.} to the radiated wave we get
\begin{subequations}
\label{init}
\beq
X^{(s)}(t\,,r)
&=&
-{\rm Re}
\left[
\sum_{n}B_ne^{-i\omega_n(t-r_*)}
\int_{-\infty}^{\infty}\frac{I(\omega\,,r){\hat X^{(s)}}_{r_+}}{A_{\rm out}}dr_*'
\right]
\label{init1}
\\
&=&
-{\rm Re}
\left[
\sum_{n}C_n e^{-i\omega_n(t-r_*)} \right]\,,
\label{init2}
\eeq
\end{subequations}
where the sum is over all poles in the complex plane and the {\it quasinormal
  excitation factors} (QNEFs) $B_n$ are defined as
\be\label{excfactors}
B_n=
\frac{A_{\rm out}}{2\omega \alpha_n}\,
\equiv
\left.
\frac{A_{\rm out}}{2\omega}
\left (\frac{dA_{\rm in}}{d\omega}\right)^{-1}
\right|_{\omega=\omega_{n}}
\,.
\ee
Here $\alpha_n$ is a commonly used notation for the derivative of $A_{\rm in}$
at the QNM frequency.  The {\it quasinormal excitation coefficients} $C_n$ can
be computed as
\be\label{exccoeffs}
C_n=B_n \int_{-\infty}^{\infty}
\frac{ I(\omega\,,r){\hat X^{(s)}}_{r_+}}{A_{\rm out}}dr_*'
\ee
whenever the integral on the right hand side, which must be evaluated at the
QNM frequency $\omega=\omega_n$, is convergent. In general the QNM frequencies
$\omega_n$, the $B_n$'s and the $C_n$'s (as well as the wavefunction) depend
on $(l,m)$ and the spin of the perturbing field $s$, but to simplify the
notation we will omit this dependence whenever there is no risk of confusion.

By definition of the QNM frequencies, ${\hat X^{(s)}}_{r_+} \sim A_{\rm
  out}e^{i\omega r_*}$ as $r_*\to \infty$ at $\omega=\omega_n$: in this sense
the above integral is ``normalized''.  For source terms $I(\omega\,,r)$ that
are zero outside some finite range of $r$, or have sufficiently rapid
exponential decay as $|r_*|\to \infty$, the integral is also convergent. For
other classes of initial data the integral as evaluated on the real line is,
in general, divergent. This is a major difference with respect to ordinary
normal mode expansions. The normal modes of closed mechanical systems are
Sturm-Liouville eigenfunctions of the wave equation, and their excitation
coefficients are weighted integrals of the source term over the mode. Since
the Sturm-Liouville eigenfunctions are always bounded, the integrals always
converge. For QNM expansions a meaningful definition of the integral for
general source terms requires more care, and the introduction of an analytical
continuation procedure \cite{lePRD,sunprice88}.

Equation (\ref{init}) is one of the main results we will use throughout the
rest of the paper. Once we specify initial data, the QNEFs allow the
determination of the QNM content of a signal. QNEFs have long been known for
scalar, electromagnetic and gravitational perturbations of Schwarzschild black
holes \cite{lePRD,sunprice88,nils95}.  More recently there have been attempts
to extend those calculations to scalar perturbations of Kerr black holes
\cite{kostasnils,quickdirty}.  Electromagnetic and (most importantly for
gravitational-wave phenomenology) gravitational perturbations of Kerr black
holes have not been dealt with so far. One purpose of the present work is to
fill this gap. Before turning to an explicit calculation of the $B_n$'s we
will provide the explicit form of the function $I(\omega\,,r)$ for $s=0$. We
will also address two important conceptual issues: the time-shift problem and
the convergence of the QNM expansion.

\subsection{Initial data for scalar perturbations}

For a scalar field \cite{kostasnils} the function $I(\omega\,,r)$ is given by
\be
I(\omega\,,r)=
e^{i\omega t_0}
\left\{
i\left [\omega-
\frac{2amr+\omega a^2 \Delta \gamma_{lm}}{(r^2+a^2)^2}\right ]
X^{(0)}(t_0\,,r)
-\left[1-\frac{a^2 \Delta \gamma_{lm}}{(r^2+a^2)^2}\right]
\dot X^{(0)}(t_0\,,r)
\right\}\,,
\label{iscal}
\ee
where $\gamma_{lm}$ is a number depending on the multipole of the field under
study:
\be \label{glm-def}
\gamma_{lm}=2\pi \int _0^{\pi}
S_{lm}^*(a\omega,\theta)S_{lm}(a\omega,\theta)
\sin^3\theta d\theta \,.\ee
Here the $S_{lm}$'s are scalar ($s=0$) spin-weighted spheroidal harmonics
\cite{BCC}. We require them to satisfy the normalization condition $2\pi
\int_0^{\pi}S_{lm}^*(a\omega,\theta) S_{lm}(a\omega,\theta)\sin\theta d\theta
=1$ (notice that this normalization differs by a factor $2\pi$ from that
adopted in \cite{kostasnils}). Even though these functions depend on
$a\omega$, in \cite{BCW} we showed that in some cases (for example, when we
are dealing with slowly damped QNMs) this dependence can safely be neglected.
When $a\omega=0$ scalar spheroidal harmonics reduce to the usual spherical
harmonics, and the integral can be computed analytically with the result
\be \label{glm-analytic}
\gamma_{lm}=\frac{2(l^2+l-1+m^2)}{(2l-1)(3+2l)}\,.
\ee
For the calculations in this paper, the analytic formula is extremely accurate
even when $a\omega\neq 0$. For example, for a near-extremal black hole with
$j=0.98$ a numerical evaluation of the integral at the fundamental scalar QNM
frequency yields $\gamma_{22}=0.854139$, in excellent agreement with the
analytic prediction $\gamma_{22}=6/7\simeq 0.857143$. Results for other
slowly-damped modes are similar.


In the following of the paper we will focus, for simplicity, on static initial
data. For large $r$ Eq.~(\ref{iscal}) reduces to
\be I(\omega\,,r)\sim i\omega e^{i\omega t_0}X^{(0)}(t_0\,,r) \,, \ee
whereas for $r$ very near the horizon we get
\be I(\omega\,,r)=ie^{i\omega t_0}\left (\omega-2m\Omega \right )X^{(0)}(t_0\,,r) \,,\ee
where $\Omega =a/r_+$ is the angular velocity of the horizon.

\subsection{The time-shift problem}

Taken at face value Eq.~(\ref{init}) is troublesome, because the response
diverges (exponentially) for early times. The well-known fact that ringdown
waveforms only make sense for certain values of $t$ is usually referred to as
the ``time-shift problem'' \cite{sunprice88,nils97,nollertprice}. For
illustration, consider Eq.~(\ref{init}) for a Schwarzschild black hole with
static initial data:
\be\label{psitr}
X^{(s)}(t\,,r)
=-{\rm Re}\left [\sum_{n=0}^{\infty}B_ne^{-i\omega_n(t-r_* - t_0)} \int_{-\infty}^{\infty}\frac{i\omega_n
X^{(s)}(t_0\,,r){\hat X^{(s)}}_{r_+}}{A_{\rm out}}dr_*'\right ] \,.
\ee

The wave field $X^{(s)}(t\,,r)$ at some observation point $r$ and time $t$
depends exponentially on the (arbitrary) choice of $t_0$, which does not make
much physical sense. Suppose in addition that the initial data consist of a
narrow pulse, say for simplicity a $\delta$-function:
$X^{(s)}(t_0,r)=\delta(r-r^S)$. Far away from the black hole $\hat
X^{(s)}_{r_+}\sim A_{\rm out} e^{i\omega_n r_*}$, and the response of the
black hole increases exponentially with $r^S$. This is against physical
intuition: according to our Fourier convention stable oscillations have ${\rm
  Im}(\omega)<0$, so a small bump in initial data far from the black hole
seems to excite much more ringing than a huge bump in the strong-field region
around the horizon.

The way out of this problem \cite{sunprice88,nils97} is to realize that
ringdown only starts after the initial data (that for simplicity are usually
assumed to have compact support and be located in the far zone, $r^S\gg 1$)
reach the potential barrier, where ringdown originates \cite{vish}, and the
perturbation travels back to the observer. Suppose the observer is sitting at
some large $r_*$ and an impulse localized at $r_*=r_*^S$ falls into the black
hole at time $t_0$. The Regge-Wheeler ``potential barrier'' has a location
depending on the mode under consideration, and the very notion of potential
barrier becomes fuzzy for Kerr black holes. A reasonable estimate is to assume
that the ``barrier'' is located close to $r_*\sim 0$, and to avoid
complications we will take this estimate as our fiducial value. In geometrical
units, based on the above discussion, the starting time for the ringdown
signal will be approximately
\be\label{tstart}
t_{\rm start}=|r_*^S|+r_*+t_0\,.
\ee
For the QNM expansion to make sense, the observation time $t$ must always be
chosen so that $t\geq t_{\rm start}$. This choice partially gets rid of the
time-shift problem, in the sense that $t_0$ and $r_*^S$ do not appear
explicitly in the waveform.

A more rigorous approach should also take into account the contribution of the
prompt response (the analogous of light cone propagation in flat space) and
the matching between these two phases. Ref.~\cite{nils97} explores some
aspects of this problem, computing the large-frequency contribution to the
Green's function (large frequency is roughly equivalent to early times). The
final result does not significantly alter our ``simple-minded'' scenario.
We should also mention that our prescription to solve the time-shift problem
works reasonably well only for sharply localized initial data, like narrow
gaussians. If the initial data have a significant spread in the radial
direction the starting time $t_{\rm start}$ is ill-defined and we must resort
to a more detailed analysis, based perhaps on time-dependent excitation
coefficients \cite{nils97}.  This problem deserves further investigation.

\subsection{Convergence of the QNM expansion}

The convergence of the QNM series has been studied in some special important
cases by Leaver \cite{lePRD}. In particular he pointed out that the sum can
only be expected to converge at times $t$ such that ringdown dominates (see
the above discussion of the time-shift problem), and that the ultimate
convergence of the series depends on the large-$n$ asymptotics of the QNEFs.
The convergence problem was revisited numerically by Andersson \cite{nils97}
computing the $B_n$'s up to $n\sim 200$ for scalar perturbations. His
calculation shows that their magnitude decreases monotonically for large $n$.
From (\ref{psitr}) it follows that, if the initial data are localized at
$r_*=r_*^S\gg 1$, the ratio of two consecutive terms behaves like
$(B_{n+1}/B_n)e^{-i(\omega_{n+1}-\omega_n)(t-r_*-r_*^S)}$.  Asymptotically,
$\omega_{n+1}\simeq \omega_n -i/2$ for large $n$.  In Appendix
\ref{app:asymptBn} we determine {\it analytically}, for the first time, the
asymptotic behavior of the QNEFs for a number of static black hole spacetimes.
Our most interesting result is that the large-$n$ behavior of scalar and
gravitational perturbations of a Schwarzschild black hole is
\be
B^{(0)}_n=B^{(-2)}_n=
-\frac{i}{3\left [\pm \log3-(2n+1)\pi i\right ]}\,,
\qquad (n\to \infty)\,,
\label{bnasbehavior}
\ee
in good agreement with numerical calculations (cf. Figure 2 in
Ref.~\cite{nils97}). This implies that $|B_{n+1}/B_n|\to 1$ in the same limit,
so the ratio of consecutive terms is of order $\exp[-(t-r_*-r_*^S)/2]$ and the
sum will converge for $(t-r_*-r_*^S)\gg 0$. This is precisely the kind of
convergence we should expect from a QNM expansion.
%

To prove more rigorously the convergence of the expansion (\ref{init}) we
still need two ingredients. The first is the convergence of the integral in
(\ref{exccoeffs}). For generic initial data this integral diverges on the real
line. The divergence can be cured by analytic continuation, evaluating the
integral as a contour integral (for details see \cite{lePRD,sunprice88}).
Secondly, the fact that the QNEFs $B_n\sim 1/n$ for large $n$ does not imply
convergence of the corresponding quasinormal excitation coefficients $C_n$. A
proof of convergence would require the calculation of the integral over
initial data {\it and} a knowledge (or an estimate of) its $n$-dependence for
large $n$.

Having discussed these basic properties of QNM expansions, in the following of
the paper we turn to an explicit calculation of the $j$-dependence of the
QNEFs for different QNMs and for different spins of the perturbing field.

\section{Quasinormal excitation factors in the Kerr background}
\label{Bn}

\subsection{Formalism}
\label{formalism}

Perturbations in the Kerr geometry can be studied using both the Teukolsky
\cite{teukolsky} and SN \cite{SN2} formalisms. There are two main advantages
in using the SN formalism: (i) the potential and source in the SN wave
equation are short ranged and (ii) in the limit $a\to 0$ the SN wave equation
reduces to the Regge-Wheeler equation, describing axial perturbations of a
Schwarzschild black hole.

The QNEFs $B_n$ depend, by their own definition (\ref{excfactors}), on the
amplitude of ingoing and outgoing waves at infinity.  Even in the
Schwarzschild limit, despite the isospectrality between axial and polar
perturbations, the Regge-Wheeler and Zerilli QNEFs differ by a proportionality
constant. Denoting quantities in the Zerilli equation by a plus (meaning
even-parity), the two are related in a simple way \cite{lePRD}:
\be
A_{\rm out}^+=
\frac{(l-1)l(l+1)(l+2)+6i\omega}{(l-1)l(l+1)(l+2)-6i\omega}
A_{\rm out}\,,\qquad
B_n^+=
\frac{(l-1)l(l+1)(l+2)+6i\omega}{(l-1)l(l+1)(l+2)-6i\omega}
B_n\,.
\ee

Similarly, the SN QNEFs for spin-$s$ perturbations $B^{(s)}$ will differ from
the Teukolsky QNEFs (that we will denote by a subscript $T$, $B^{(s)}_{\rm
  T}$) by a proportionality constant. In Appendix \ref{app:SN} we derive the
following relations:
\be\label{Bn-alls}
B^{(0)}_{\rm T}=B^{(0)}\,,
\qquad
B^{(-1)}_{\rm T}=-\frac{4\omega^2}{2am\omega-A_{lm}-a^2\omega^2}B^{(-1)}\,,
\qquad
B^{(-2)}_{\rm T}=
\frac{16\omega^4}
{\lambda(\lambda+2)-6i\omega-12a\omega(a\omega-m)}
B^{(-2)}\,.
\ee
with $\lambda\equiv A_{lm}+(a\omega)^2-2am\omega$.

Our calculations in this paper rely heavily on the formalism developed by
Leaver \cite{lePRS,lePRD,leJMP}, which refers to the Teukolsky formulation of
the Kerr perturbation equations.  In this Section we only outline the basic
steps of the calculation, relegating details to Appendix \ref{app:tech}.
Expanding a spin-$s$ field $\psi(t\,,r\,,\theta\,,\phi)$ as
\be
\psi(t\,,r\,,\theta\,,\phi)
=\frac{1}{2\pi}\int e^{-i\omega t} \sum_{l=|s|}^{\infty}\sum_{m=-l}^{l}
e^{im\phi}S_{lm}(\theta)R_{lm}(r)d\omega\,,
\ee
we get ordinary differential equations for $S_{lm}$ and $R_{lm}$
\cite{teukolsky}:
\beq
& & \partial_u \left [(1-u^2)\partial_u S_{lm}\right ]+\left (a^2\omega^2u^2-2a\omega s u +s+A_{lm}
-\frac{(m+s u)^2}{1-u^2}  \right )S_{lm}=0\,,\\
& & \Delta \partial^2_r R_{lm}+(s+1)(2r-1)\partial_rR_{lm}+VR_{lm}=0\,,
\label{radial}
\eeq
where $u\equiv \cos\theta$ and
\be
V=2is\omega r -a^2\omega^2-A_{lm}+\frac{1}{\Delta}\left [(r^2+a^2)^2\omega^2-2am\omega r+
a^2m^2+is\left(am(2r-1)-\omega(r^2-a^2)        \right )   \right ]\,.
\ee
The field spin $s=0\,,-1\,,-2$ for scalar, electromagnetic and gravitational
perturbations, respectively. The separation constant $A_{lm}$ reduces to
$l(l+1)-s(s+1)$ in the Schwarzschild limit. The general series solution of the
angular equation with appropriate boundary conditions is given in Appendix
\ref{app:angular}.

Let us consider the radial equation (\ref{radial}). For brevity, in the rest
of the paper we drop the dependence of the Teukolsky function $R_{lm}$ on the
angular indices $(l,m)$. Following Leaver (see Appendix \ref{app:tech} for
details) we introduce three different solutions of the homogeneous equation
(\ref{radial}): $R_{r_+}\,,R_{\infty_+}\,,R_{\infty_-}$, with asymptotic
behavior \cite{lePRD}
\begin{subequations}
\label{teuk-norm}
\beq
& & \lim_{r \to r_+}R_{r_+}\sim
(r_+-r_-)^{-1-s+i\omega+i\sigma_+}
e^{i\omega r_+}(r-r_+)^{-s-i\sigma_+}\,,
\label{Rrp-norm}
\\
& & \lim_{r \to \infty }R_{r_+}\sim A_{\rm in}^T(\omega) r^{-1-i\omega} e^{-i\omega r}+A_{\rm
out}^T(\omega)r^{-1-2s+i\omega}e^{i\omega r}\,,
\label{psirmainf}
\eeq
\end{subequations}
\begin{subequations}
\beq
&&\lim_{r \to \infty }R_{\infty_+}\sim r^{-1-2s+i\omega}e^{i\omega r}\,, \\
&&\lim_{r \to \infty }R_{\infty_-}\sim r^{-1-i\omega}e^{-i\omega r}\,.
\label{asymp22}
\eeq
\end{subequations}
In these relations $\sigma_+=\left(\omega r_+-am\right)/b$ and
$b=\sqrt{1-4a^2}$.

By definition, QNM frequencies $\omega_{n}$ are such that $A_{\rm
  in}^T(\omega_{n})=0$. We compute them following the procedure in
\cite{lePRS,cardosoberti} and verifying that $A_{\rm in}^T=0$ to some required
numerical accuracy (typically better than one part in $10^{6}$).  From
(\ref{psirmainf}) and (\ref{asymp22}) it follows that
\begin{subequations}
\beq
R_{r_+}&=&A_{\rm in}^TR_{\infty_-}+A_{\rm out }^TR_{\infty_+}\,,\\
R_{r_+}'&=&A_{\rm in}^TR_{\infty_-}'+A_{\rm out
}^TR_{\infty_+}'\,.
\eeq
\label{match}
\end{subequations}
From these relations it is clear that computing the QNEFs requires accurate
representations of the solutions both close to the horizon and at infinity.
Close to the horizon, an accurate representation is provided by a series
solution first found by Jaff\'e. At infinity we must resort to a different
expansion in terms of Coulomb wavefunctions. Properties of these
representations which are important for the calculation of the three solutions
($R_{r_+}$, $R_{\infty_+}$ and $R_{\infty_-}$) are given in Appendix
\ref{app:jaffe} and \ref{app:coulomb}, respectively. We performed a number of
consistency checks on the solutions, some of which are described in Appendix
\ref{app:checks}. More details can be found in the original papers by Leaver
\cite{lePRS,lePRD,leJMP}. Notice that the outgoing-wave amplitude $A_{\rm
  out}$ is unambiguously defined once we fix the normalization according to
Eq.~(\ref{teuk-norm}). The corresponding normalization for the SN
wavefunctions is derived in Appendix \ref{app:SN}: see in particular
Eqs.~(\ref{normsn-s0}), (\ref{normsn-s1}) and (\ref{normsn-s2}).

To compute the QNEFs (\ref{Bn-alls}) as functions of the rotation rate $j=2a$,
we first obtain the amplitudes $A_{\rm in}^T$ and $A_{\rm out}^T$ by matching
the three solutions and their derivatives at some point (usually $r=5-7$).
Then we compute the derivative $\alpha_n$ introduced in (\ref{excfactors}).
Close to a QNM frequency we can perform a Taylor expansion of the ingoing
amplitude: $A_{\rm in}^T\simeq \alpha_n \delta \omega$, with $\delta \omega$ a
small complex number. For each value of $s$ and $a$ we evaluate $A_{\rm in}^T$
at a discrete set of points $\omega=\omega_n+k\delta \omega$ (typically we
pick $\delta \omega=10^{-4}$ and the integer $k=-2,\dots,2$). Then we obtain
$\alpha_n$ from a linear fit of $A_{\rm in}^T$ as a function of $\delta
\omega$, and verify that the resulting derivatives satisfy the Cauchy-Riemann
conditions.

In the rest of this paper we present our results for the QNEFs using both the
Teukolsky and SN formalisms, and discuss some of their implications for
gravitational-wave phenomenology.

\subsection{Numerical results}
\label{numerics}

The QNEFs of Schwarzschild black holes were first introduced and computed by
Leaver in his seminal analysis of the radiative Green's function. Table I of
\cite{lePRD} lists the QNEFs for the first four overtones of electromagnetic
perturbations with $l=1$, the first seven overtones of gravitational
perturbations with $l=2$ and the first four overtones of gravitational
perturbations with $l=3$ and $l=4$.  Building on Leaver's analysis, Sun and
Price \cite{sunprice88} studied the initial data dependence of the
Schwarzschild quasinormal excitation {\it coefficients}. Andersson
\cite{nils95} computed the QNEFs for scalar and gravitational perturbations
using the approximate phase-integral method, finding good quantitative
agreement with Leaver for $s=-2$ and pointing out a few sign errors in
Leaver's results. Later, using the same technique he was able to compute the
(scalar) QNEFs up to overtone numbers $n\sim 200$ \cite{nils97}. Glampedakis
and Andersson \cite{quickdirty} tried to extend the methods of
Ref.~\cite{nils95} to scalar perturbations of Kerr black holes. The QNEFs
listed in their Table 3 do not reduce to the correct Schwarzschild limit
\cite{noteqd}, and the ``effective amplitude'' results in Figure 5 of their
paper are only qualitatively correct.

We computed QNEFs using both Mathematica and a Fortran code. In the limit
$a\to 0$ the Fortran code reproduces Table I of \cite{lePRD}, once we correct
for Leaver's sign mistakes, to all (five) significant digits. Given the
complexity of the algorithm described in Appendix \ref{app:tech} it is hard to
quantify our numerical error for $a\neq 0$, but results should be accurate to
the same level.  The numerical accuracy depends on many factors, including the
number of terms included in the Coulomb wavefunction expansion (\ref{final}),
the matching radius chosen to solve the linear system (\ref{match}) and the
value of $\delta \omega$ used to compute $\alpha_n$ by a linear fit.  In
general, the number of digits to which $A_{\rm in}=0$ at the QNM frequency and
the level to which the Cauchy-Riemann conditions are satisfied in the
calculation of $\alpha_n$ are good accuracy indicators. We decided to list the
QNEFs with a five-digit accuracy, but in some cases (especially for corotating
modes with $l=m$ and large rotation rates) the number of significant digits
may be smaller.

\subsubsection{Scalar perturbations}

In Tables \ref{tab:omegas0} and \ref{tab:Alms0} we provide the QNM frequencies
and damping times for scalar perturbations with $l=2$, all values of $m$ and
selected values of the angular momentum $j=2a$.  In Table
\ref{tab:scalarexccoef} we list the corresponding QNEFs, and in Table
\ref{tab:scalarexccoef2} the outgoing-wave amplitudes $A_{\rm out}$. In the
Schwarzschild limit, our QNEFs can be compared with Andersson's
\cite{nils95,nils97}.
His result for the fundamental scalar mode with $l=2$ agrees with ours to a
five-digit accuracy \cite{notenils95}. This agreement is quite impressive,
given the approximate nature of the phase-integral method. For $l=0$ his
result for the fundamental mode ($B^{(0)}_0\simeq 0.197-0.046i$) is within
$\sim 7 \%$ ($\sim 23 \%$) of the real (imaginary) part of our result:
$B^{(0)}_0\simeq 0.212-0.059i$. The agreement is still very good, considering
that the phase-integral method is based on a WKB-type approximation and should
only be accurate in the ``eikonal'' (large $l$) limit.

\begin{figure*}[ht]
\begin{center}
\begin{tabular}{cc}
\epsfig{file=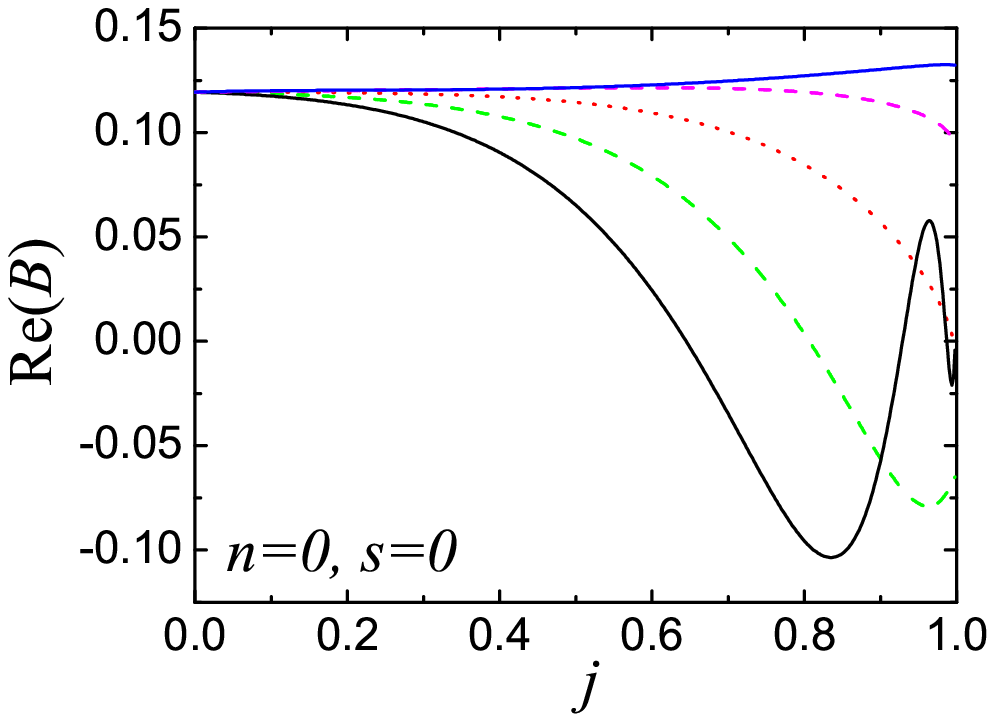,width=9cm,angle=0} &
\epsfig{file=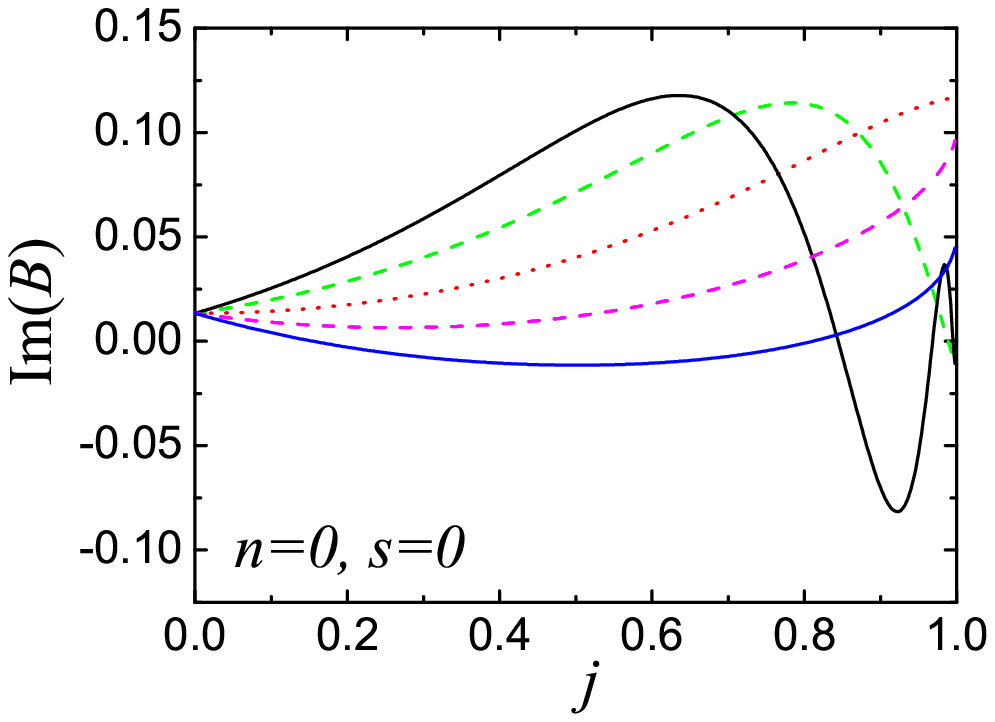,width=9cm,angle=0} \\
\end{tabular}
\end{center}
\caption{Real part (left) and imaginary part (right) of the scalar QNEFs for
  the fundamental mode with $l=2$. Solid lines correspond to $|m|=2$, dashed
  lines to $|m|=1$, the dotted line to $m=0$. The lines displaying
  oscillations for large $j$ (black and green in the colour version) have
  $m>0$.}
\label{fig:Bns0ri}
\end{figure*}

Some physical insight can be obtained by plotting the real and imaginary parts
of the scalar QNEFs for the fundamental mode with $l=2$ as functions of $j$
(Fig.~\ref{fig:Bns0ri}). The excitation of modes with $m\leq 0$ is a
slowly-varying function of the rotation rate. On the contrary, as $j\to 1$ the
real and imaginary parts of $B^{(0)}$ for corotating modes become rapidly
oscillating functions of $j$. Our calculation is in remarkable agreement with
a classic result by Ferrari and Mashhoon \cite{poschl}: when $l=m$ the QNM
excitation tends to zero as $j\to 1$, and this is an indication (if not a
proof, given the incompleteness of QNMs) that extremal Kerr black holes are
{\it not} marginally unstable.

\begin{figure*}[ht]
\begin{center}
\begin{tabular}{cc}
\epsfig{file=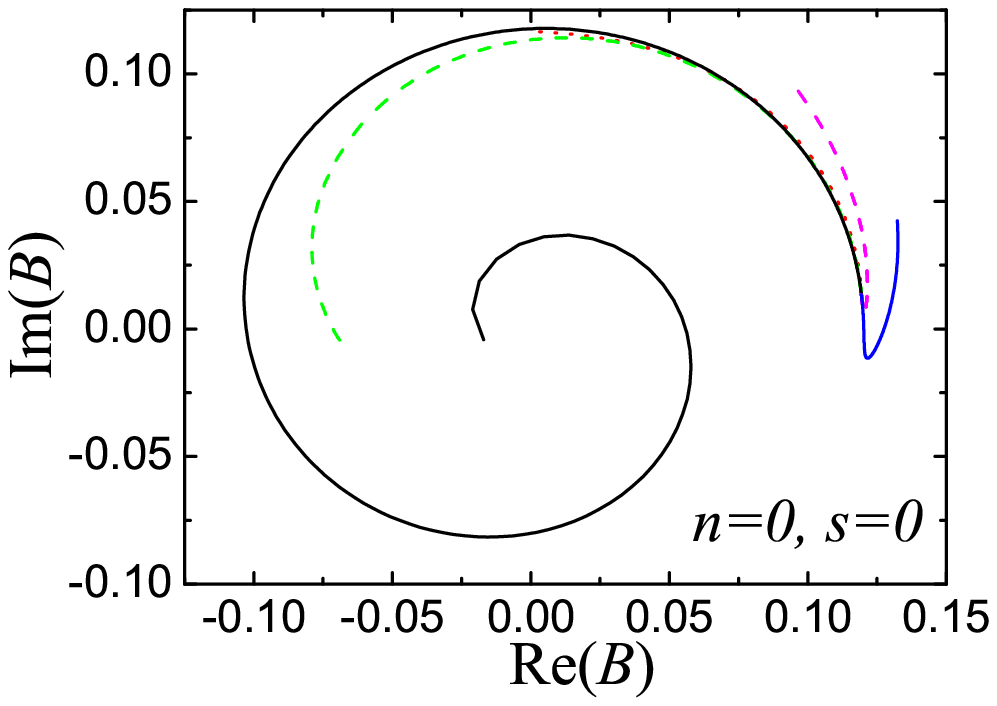,width=9cm,angle=0} &
\epsfig{file=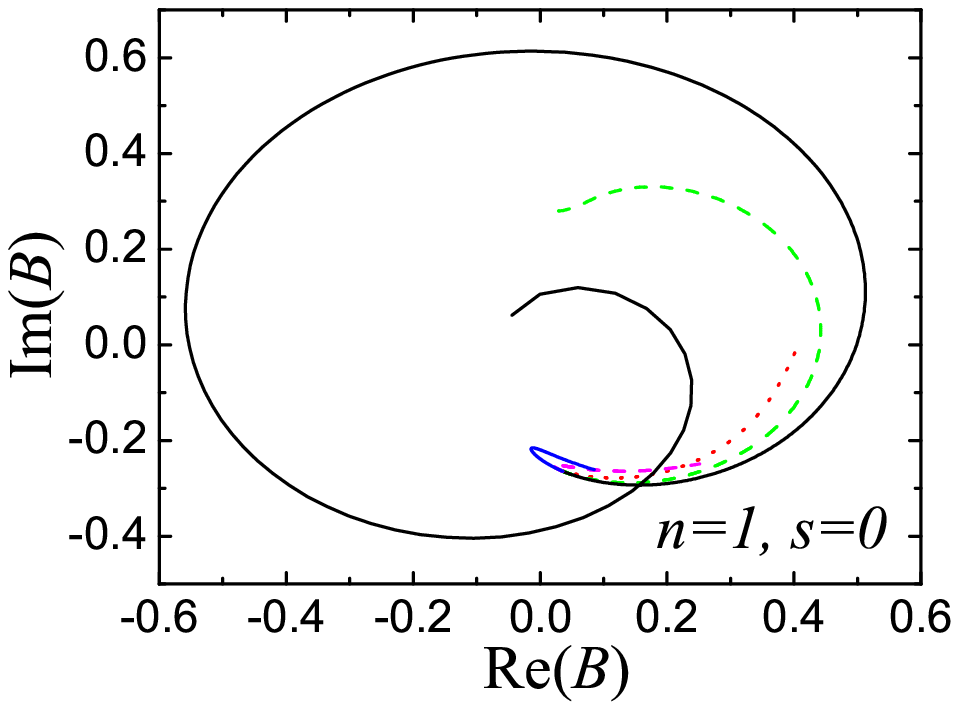,width=9cm,angle=0} \\
\end{tabular}
\end{center}
\caption{Path of the scalar QNEFs for the fundamental mode (left) and first
  overtone (right) with $l=2$ as $j$ varies in the range $0\leq j\leq 0.996$.
  Linestyles are the same as in Fig.~\ref{fig:Bns0ri}.}
\label{fig:Bns0}
\end{figure*}

More features can be seen in Fig.~\ref{fig:Bns0}, where we plot the path
followed by the real and imaginary parts of $B^{(0)}_n$, thought of as
parametric functions of $j$, for $0\leq j\leq 0.996$. The oscillations of the
real and imaginary parts of $B^{(0)}$ for the corotating mode with $l=m=2$
produce a spiral in the complex plane. For small $j$, the QNEFs of corotating
and counterrotating modes with the same $|m|$ move in opposite directions,
approaching the Schwarzschild limit with the same tangent. Corotating and
counterrotating QNM frequencies tend to their Schwarzschild limit in a similar
way \cite{bertiqnm,cardosoberti}. Another noteworthy feature of
Fig.~\ref{fig:Bns0} is the scale of the real and imaginary axes: for
corotating modes with $l=m$ and large rotation, $|B^{(0)}_1|$ is roughly five
times larger than $|B^{(0)}_0|$. In other words, {\it for perturbations with
  $l=m$ the contribution of the overtones should be more significant for
  rapidly rotating black holes.} In the following we will see that many of
these considerations are still valid when we consider perturbations of spin
$s\neq 0$.

\subsubsection{Electromagnetic perturbations}


It is widely believed that astrophysical black holes should possess very
little charge (if any).  For this reason electromagnetic perturbations of a
Kerr black hole ($s=-1$) are not considered of great astrophysical relevance.
Nonetheless, electromagnetic perturbations are of more than academic interest.
For example, they could find useful applications in models of elementary
particles based on the Kerr-Newman metric (see \cite{BK-KN} and references
therein).

\begin{figure*}[ht]
\begin{center}
\begin{tabular}{cc}
\epsfig{file=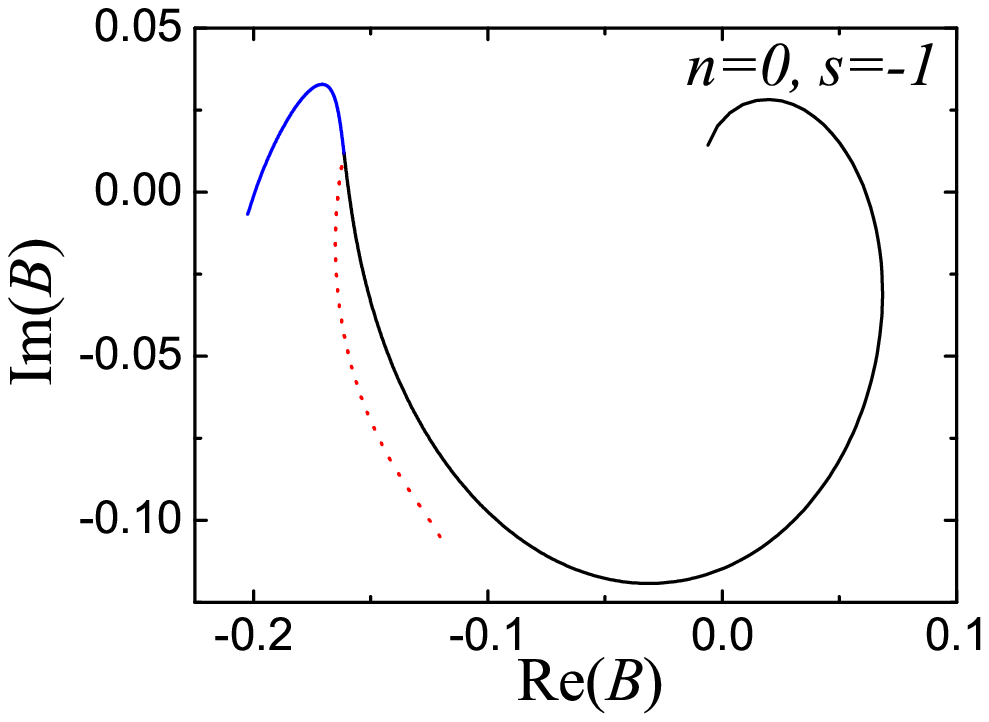,width=9cm,angle=0} &
\epsfig{file=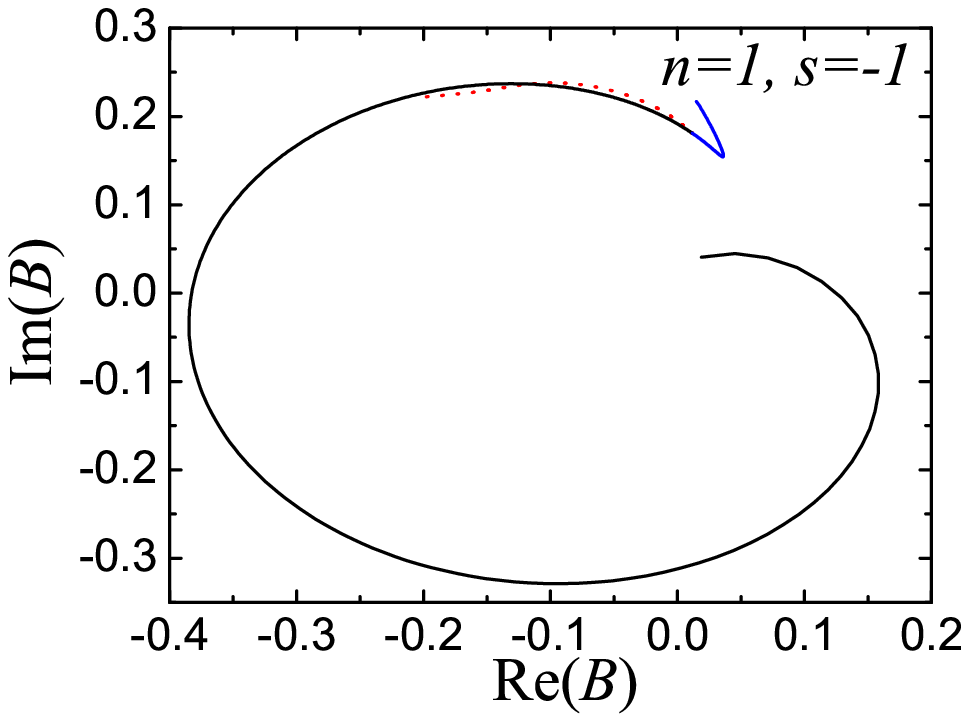,width=9cm,angle=0} \\
\epsfig{file=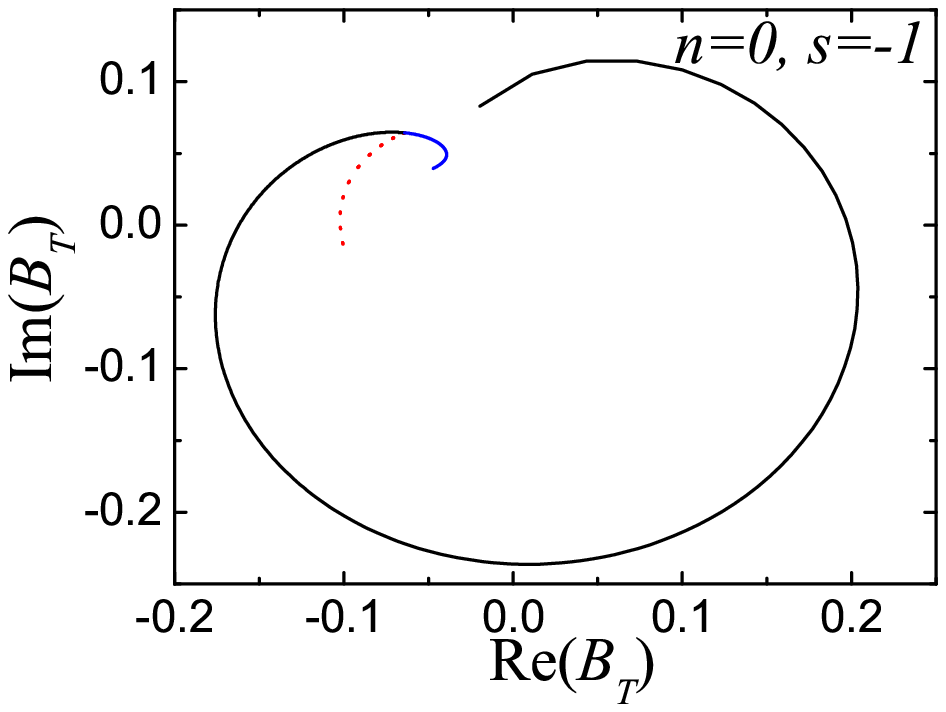,width=9cm,angle=0} &
\epsfig{file=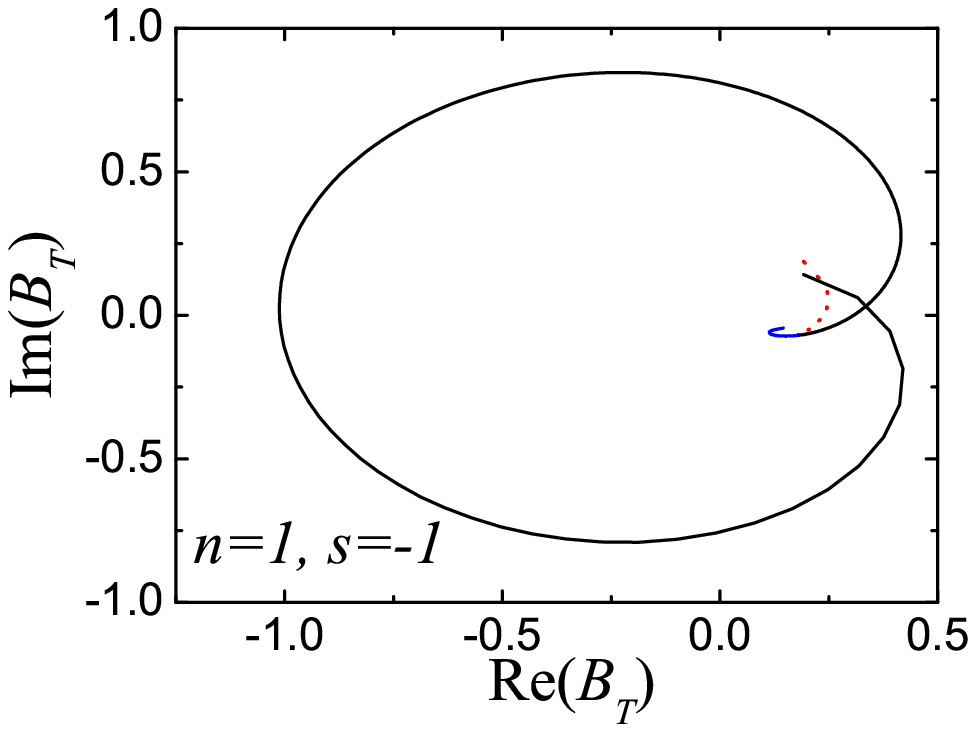,width=9cm,angle=0} \\
\end{tabular}
\end{center}
\caption{Path of the electromagnetic QNEFs for the fundamental mode (left) and
  first overtone (right) with $l=1$ as $j$ varies in the range $0\leq j\leq
  0.996$. The top panels refer to the SN formalism, the bottom panels to the
  Teukolsky formalism. Linestyles are the same as in Fig.~\ref{fig:Bns0ri}.}
\label{fig:Bns1}
\end{figure*}

Results for the lowest radiative multipole $l=1$ are presented in
Fig.~\ref{fig:Bns1}. Compared with the scalar case, the main new feature is
that QNEFs differ (both analytically and numerically) depending on whether we
use the SN or Teukolsky formalisms: see Eq.~(\ref{Bn-alls}) and Appendix
\ref{app:SN}. Of course, both formalisms are equally legitimate, and the use
of one or the other depends on the physical problem at hand. By construction
the SN QNEFs reduce to the Regge-Wheeler QNEFs in the Schwarzschild limit, and
in this sense they have a more direct physical interpretation. As $j\to 0$ the
(Bardeen-Press-)Teukolsky quantities can be transformed to the corresponding
metric quantities in the Zerilli (Regge-Wheeler) formalism using the
differential transformations derived by Chandrasekhar \cite{MTB}.

Apart from this distinction, the qualitative features of Fig.~\ref{fig:Bns1}
are similar to the scalar case of Fig.~\ref{fig:Bns0}. For small $j$, the
QNEFs of corotating and counterrotating modes with the same $|m|$ approach the
Schwarzschild limit with the same tangent. For corotating modes with $l=m=1$
and large rotation rates, $|B^{(-1)}_1|$ is significantly larger than
$|B^{(-1)}_0|$.  This is another indication that, independently of the the
value of $s$, the high-overtone contribution should be more significant for
$l=m$ modes and rapidly rotating black holes.

\subsubsection{Gravitational perturbations}

Gravitational QNEFs in the Teukolsky and SN formalisms are listed in Tables
\ref{tab:gravexccoef}-\ref{tab:gravexccoef2} for the first two overtones with
$l=2$, and selected values of the black hole's angular momentum. The
corresponding QNM frequencies and angular separation constants can be found in
Tables II and V of \cite{BCW}, respectively (there we list the imaginary parts
with the opposite sign). As $j\to 0$ our numbers agree to all digits with
Table I of \cite{lePRD}, except for the relative sign of a few modes with
$l=3$ and $l=4$. These minor sign errors in Leaver's paper were first pointed
out in \cite{nils95}.

\begin{figure*}[ht]
\begin{center}
\begin{tabular}{cc}
\epsfig{file=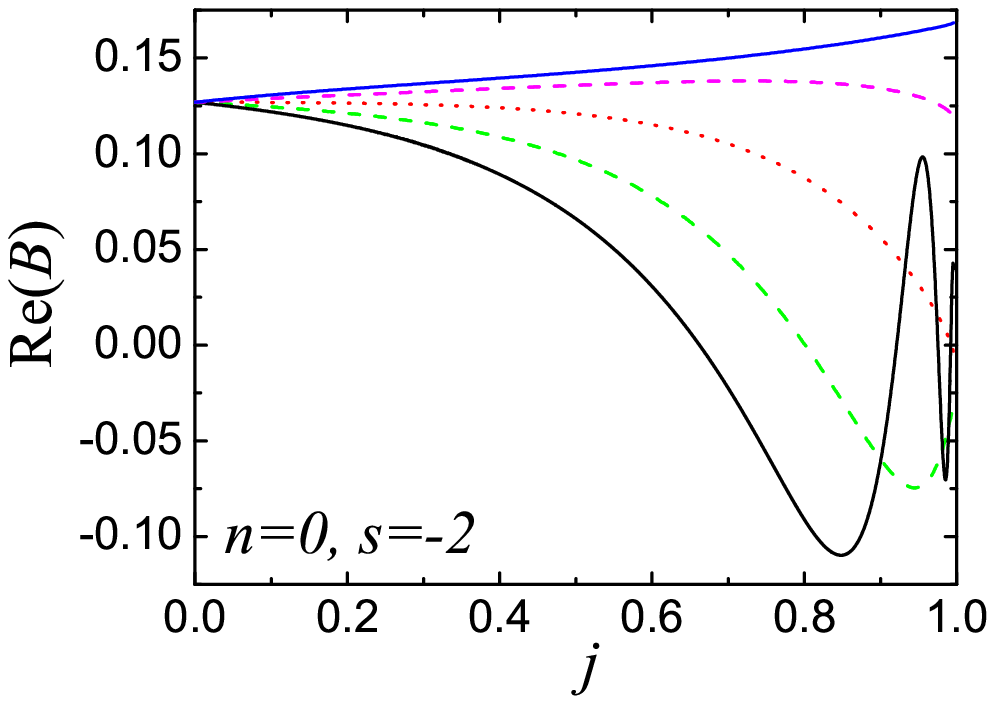,width=9cm,angle=0} &
\epsfig{file=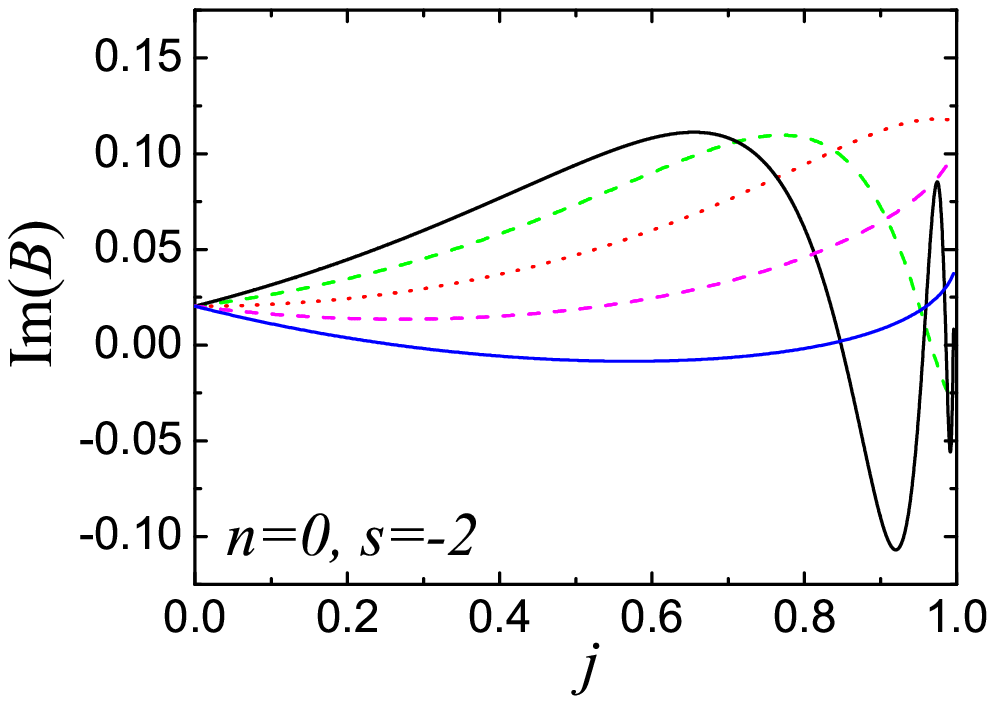,width=9cm,angle=0} \\
\epsfig{file=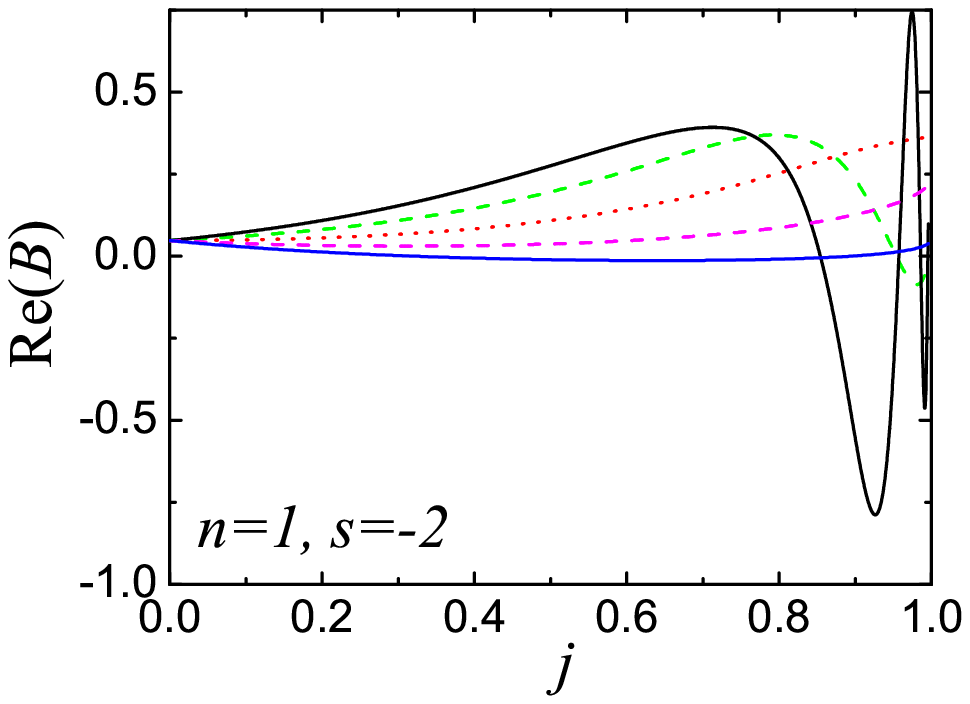,width=9cm,angle=0} &
\epsfig{file=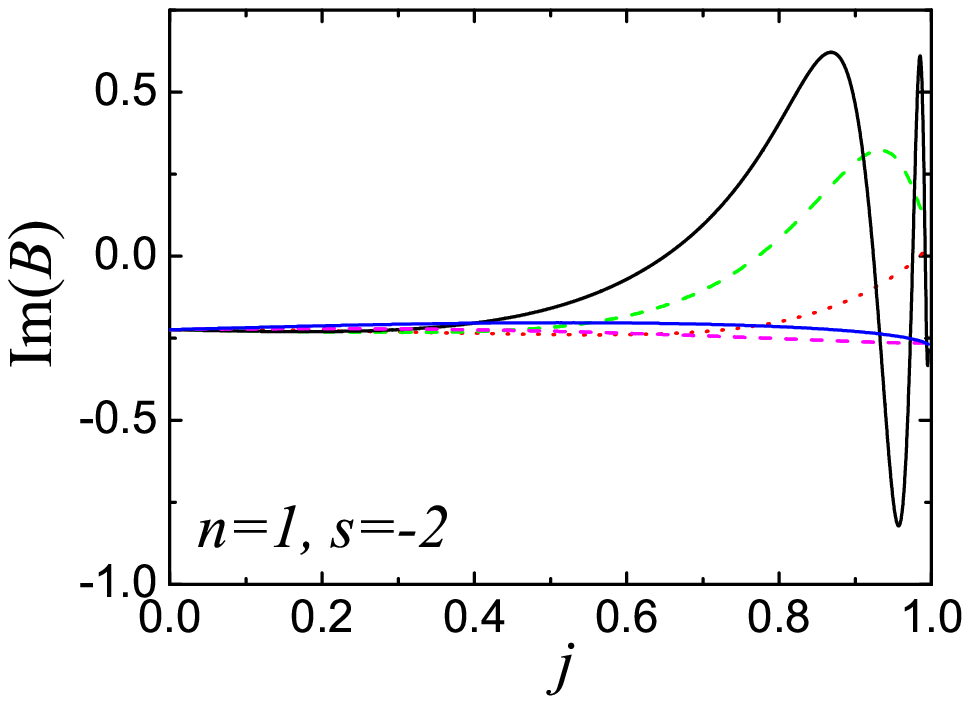,width=9cm,angle=0} \\
\end{tabular}
\end{center}
\caption{Real part (left) and imaginary part (right) of the gravitational
  QNEFs with $l=2$ and different values of the overtone index $n$, indicated
  in the inset. All plots refer to the SN formalism. Linestyles are the same
  as in Fig.~\ref{fig:Bns0ri}.}
\label{fig:Bns2ri}
\end{figure*}

In Fig.~\ref{fig:Bns2ri} we plot the real and imaginary parts of the
gravitational QNEFs for the first two overtones with $l=2$ in the SN formalism
as functions of $j$. The qualitative behavior is remarkably similar to the
scalar case (cf. Fig.~\ref{fig:Bns0ri}).  For fast rotation, the maximum
excitation of the first gravitational overtone (as compared to the maximum
excitation of the fundamental mode) is larger by a factor $\sim 8$.

\begin{figure*}[ht]
\begin{center}
\begin{tabular}{cc}
\epsfig{file=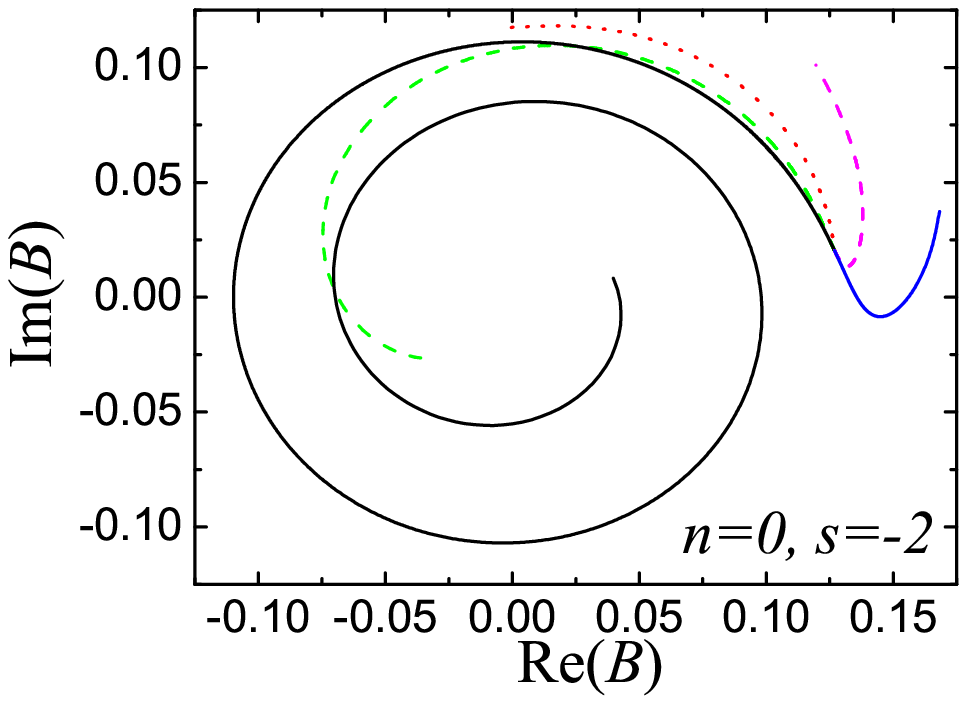,width=9cm,angle=0} &
\epsfig{file=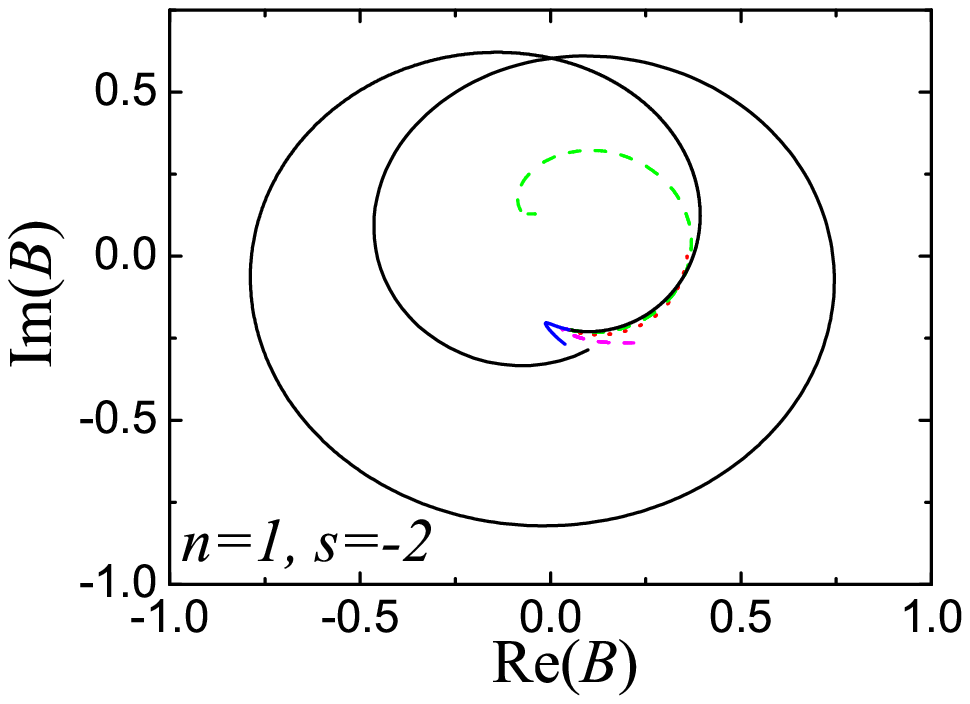,width=9cm,angle=0} \\
\epsfig{file=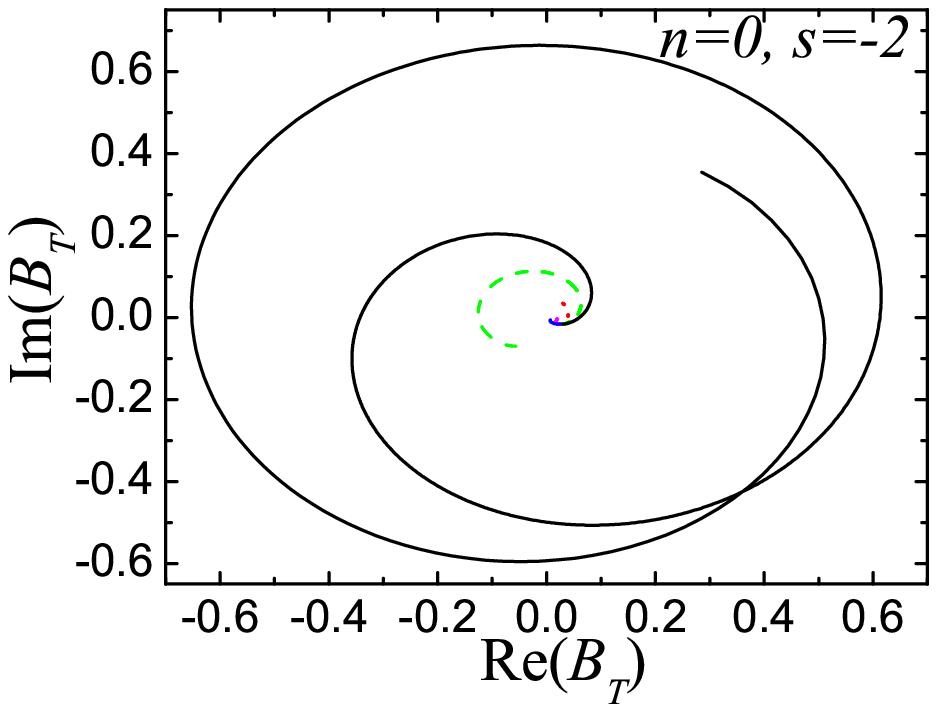,width=9cm,angle=0} &
\epsfig{file=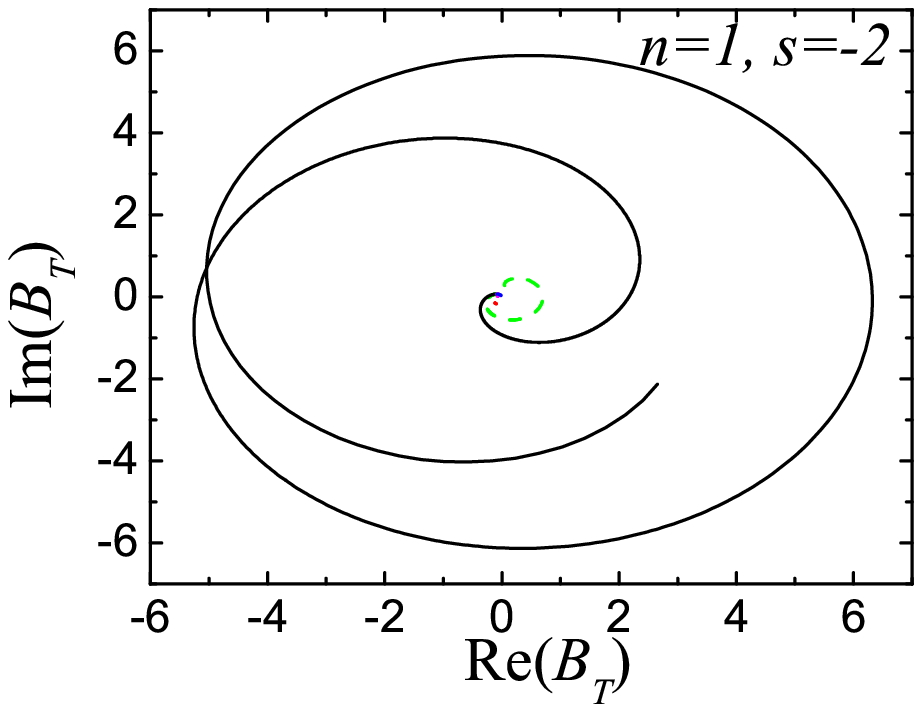,width=9cm,angle=0} \\
\end{tabular}
\end{center}
\caption{Path of the gravitational QNEFs for the fundamental mode (left) and
  first overtone (right) with $l=2$ as $j$ varies in the range $0\leq j\leq
  0.996$. The top panels refer to the SN formalism, the bottom panels to the
  Teukolsky formalism. Linestyles are the same as in Fig.~\ref{fig:Bns0ri}.}
\label{fig:Bns2}
\end{figure*}

In Fig.~\ref{fig:Bns2} we plot the path followed by the real and imaginary
parts of $B^{(-2)}_n$, thought of as parametric functions of $j$, in the range
$0\leq j\leq 0.996$. There's a remarkable similarity between the plot for
$n=0$ in the SN formalism and the scalar case of Fig.~\ref{fig:Bns0}. Even
though it's hard to accurately compute the QNEFs when $j$ is very close to
one, our numerics show clear evidence that in both cases the excitation of
$l=m$ modes is zero in this limit: the center of the spiral is located at the
origin of the complex plane. Our calculation provides further evidence that
extremal Kerr black holes are stable, and that QNMs of a fast rotating Kerr
black hole are {\it very hard to excite} \cite{poschl,kostasnils,quickdirty}.

\begin{figure*}[ht]
\begin{center}
\begin{tabular}{cc}
\epsfig{file=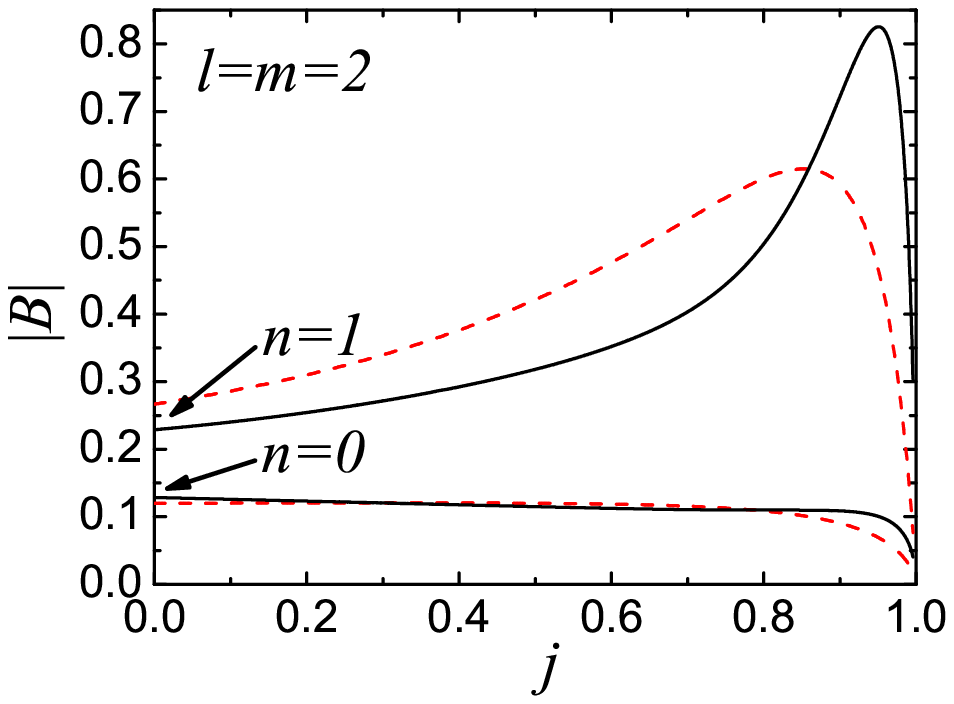,width=9cm,angle=0}
\end{tabular}
\end{center}
\caption{Modulus of the SN QNEFs for the fundamental mode and first overtone
  with $l=m=2$. Solid (black) lines are for gravitational perturbations,
  dashed (red) lines for scalar perturbations.}
\label{fig:plot-mod}
\end{figure*}

In Fig.~\ref{fig:plot-mod} we plot the modulus of gravitational and scalar
QNEFs for the fundamental mode and first overtone with $l=m=2$. This plot is
useful to visualize the effect of rotation on the excitation of different
overtones, at least for radially localized initial data (for gaussian initial
data the response is exponentially modulated by the QNM frequencies, see
below). For gravitational perturbations the first overtone is maximally
excited (with a relative amplification of a factor $\sim 8$ with respect to
the fundamental mode) for $j\simeq 0.95$. For scalar perturbations the effect
is similar, but the relative amplification of the first overtone is smaller
(roughly a factor $6$) and the maximum occurs at $j\simeq 0.85$.

\subsection{A simple application: localized and gaussian initial data}

As an application of our calculation of the QNEFs, we study the response of a
Kerr black hole to initial data in simple situations amenable to an analytic
treatment. The relevance of these model problems to realistic perturbations of
astrophysical black holes is questionable, but our analysis could provide some
insight into the black hole's response to more generic initial data. For
simplicity in the following discussion we focus on scalar perturbations, so
that Eq.~(\ref{iscal}) applies. We start by considering static, localized
initial data of the form
\be
X^{(0)}(t_0\,,r)=\delta (r_*-r_*^S)\,, \qquad
\dot X^{(0)}(t_0\,,r)=0  \,.
\ee
Combining Eq.~(\ref{init}) with the definition (\ref{tstart}) yields
\be
X^{(0)}(t\,,r)=
-{\rm Re}\left [\sum_{n=0}^{\infty}i B_n
e^{-i\omega_n(t-t_{\rm start})}
\frac{{\hat X^{(0)}}_{r_+}(r_*^S)e^{-i\omega_n |r_*^S|}}{A_{\rm out}}
\left.\left(\omega_n-\frac{2amr+\omega_n
a^2\gamma_{lm}\Delta}{(r^2+a^2)^2} \right)\right|_{r=r^S} \right ] \,.
\label{pointbhexact}
\ee

For initial data in the far zone this is further simplified to
\be X^{(0)}(t\,,r)\simeq
-{\rm Re}
\left[\sum_{n=0}^{\infty} i\omega_n B_ne^{-i\omega_n(t-t_{\rm start})}\right]
\,,\quad r_*^S\rightarrow +\infty \,.\label{pointsourcefaraway}
\ee
For initial data localized near the horizon, using (\ref{asrplus}) we
get
\be X^{(0)}(t\,,r)\simeq
-{\rm Re}
\left[\sum_{n=0}^{\infty}i
\left(\omega_n -2m\Omega\right )\frac{B_n}{A_{\rm out}}
e^{-i\omega_n(t-t_{\rm start})}
e^{im\Omega r_*^S} \right]
\,,\quad r_*^S \rightarrow -\infty\,.
\label{pointsourceclose}
\ee
Rotational effects appear (in the form of the usual Lense-Thirring frame
dragging) only when the $\delta$-like disturbance is located near the horizon.
Eq.~(\ref{bnasbehavior}) shows that in the Schwarzschild case $B_n \sim 1/n$
for large $n$. Since we expect the same leading-order behavior to hold for
Kerr black holes, and we know that Kerr QNM frequencies scale as $\omega_n\sim
-in$ for large $n$ \cite{cardosoberti}, a $\delta$-like source excites all
modes to comparable amplitude. This is reminiscent of the analogous result
(\ref{pointstring}) for the vibrating string.

The vibrating string analogy suggests that we may be able to associate nodes
to the QNM eigenfunctions. In other words, there could be special locations of
the initial data $r_*^S$ for which the black hole's response is zero (or at
least very small). We expect it should be easier to find these ``nodes'', if
they exist at all, for QNM eigenfunctions which are as close as possible to
ordinary normal mode eigenfunctions. This happens when ${\rm Re}(\omega_n)\gg
{\rm Im}(\omega_n)$, that is, for corotating modes with $l=m$ and
near-extremal ($j\simeq 1$) black holes \cite{bertiqnm}. For this reason below
we focus on scalar perturbations of Kerr black holes with $j=0.98$ and
consider only modes with $l=m=1$ or $l=m=2$.

\begin{figure*}[ht]
\begin{center}
\begin{tabular}{cc}
\epsfig{file=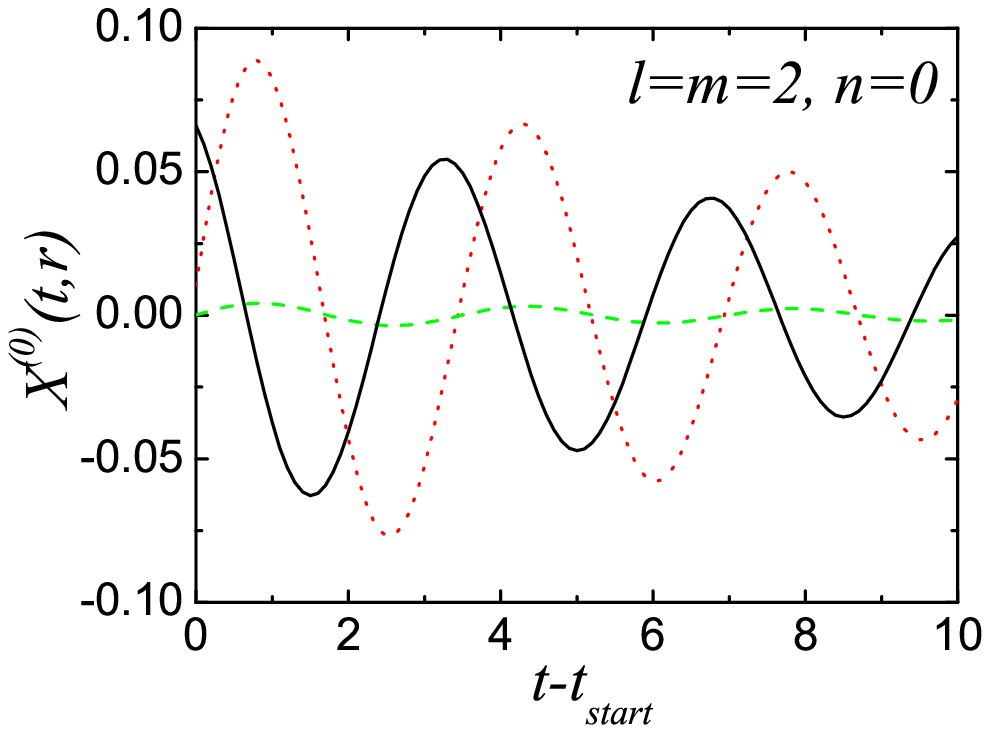,width=9cm,angle=0} &
\epsfig{file=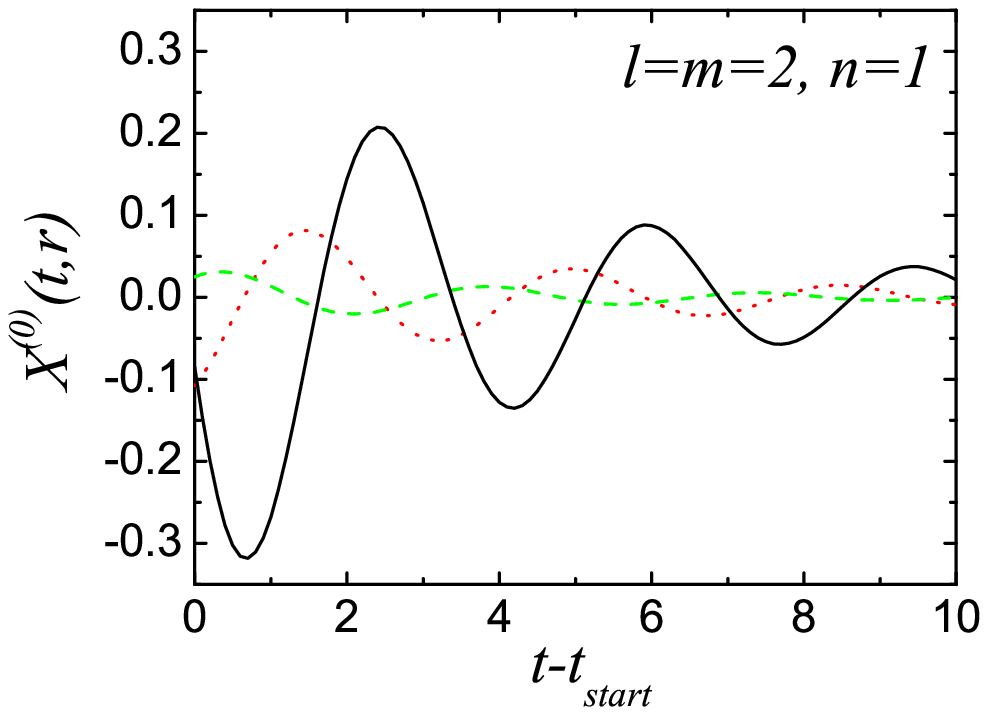,width=9cm,angle=0} \\
\end{tabular}
\end{center}
\caption{Time-domain single-mode waveforms for scalar perturbations of a Kerr
  black hole with $j=0.98$. Each panel shows results for different values of
  $(l,m,n)$. Dotted (red) lines correspond to initial data located extremely
  close to the horizon, at $r^S=0.6$.  Solid (black) lines are obtained for
  initial data localized relatively far away from the black hole, at $r^S=2$.
  Dashed (green) lines show the response for initial data localized at the
  point of minimal excitation, that we determined to be:
%
$r^S\simeq 0.86$ ($l=m=2$, $n=0$);
$r^S\simeq 0.87$ ($l=m=2$, $n=1$).
} \label{fig:pluck}
\end{figure*}

In Fig.~\ref{fig:pluck} we plot the response of a Kerr black hole with
$j=0.98$ (so that the horizon radius $r_+ \sim 0.5994$) to initial data
localized at different values of $r^S$, using Eq.~(\ref{pointbhexact}). We
compute the QNM frequency $\omega_n$ and the angular eigenvalue $A_{lm}$ in
(\ref{pointbhexact}) using Leaver's method \cite{lePRS}.  The eigenfunction
${\hat X^{(0)}}_{r_+}(r_*^S)$ can be obtained combining Eq.~(\ref{Rrmaisdet})
and (\ref{TSNs0}).  The values of $B_n$ and $A_{\rm out}$ can be read off
Tables \ref{tab:scalarexccoef} and \ref{tab:scalarexccoef2}, respectively.
For increased accuracy we evaluate the $\gamma_{lm}$'s numerically from their
definition (\ref{glm-def}), integrating the (complex) angular QNM
eigenfunctions (see Appendix \ref{app:angular}), but the numerical results
differ only marginally from the simple analytic approximation
(\ref{glm-analytic}).

For fixed $(l,m)$ and $n=0$ the black hole's response has a relative maximum
when the initial data are located very close to the horizon. As $r^S$
increases the wave amplitude decreases monotonically, reaching a minimum at
$r^S\simeq 0.86$ (the exact location is only weakly sensitive to the
particular mode we consider). This radial location can be seen as
corresponding to a ``node'' of the QNM eigenfunction. We also show wave
amplitudes for initial data located relatively far from the horizon, at
$r^S=2$. These amplitudes are already quite close to the amplitudes for
initial data ``falling from infinity''.

The association of nodes with the eigenfunctions is more problematic and
ambiguous for larger damping. For example, if we consider the fundamental mode
of a Schwarzschild black hole (for which ${\rm Re}(\omega_n)$ and ${\rm
  Im}(\omega_n)$ are comparable) the wave amplitude turns out to be only
weakly sensitive to the location $r^S$ of initial data. The same happens when
we consider counterrotating modes of (slowly or rapidly rotating) Kerr black
holes.

A natural next step is to study non-localized initial data. We consider static
initial data given by a gaussian wavepacket
\be
X^{(0)}(t_0\,,r)= k e^{-b(r_*-r_*^S)^2} \,,
\ee
where $k$ and $b$ are arbitrary constants.  For Schwarzschild black holes, the
scattering of gaussian initial data was first studied numerically in the time
domain by Vishveshwara \cite{vish}. An interpretation of the results in terms
of QNM expansions for initial data in the far zone was later provided by
Andersson \cite{nils95}.  Following \cite{nils95}, for initial data in the far
zone (large $r_*^S$) we get
\beq
X^{(0)}(t\,,r)&\simeq& -{\rm Re}
\left [\sum_{n=0}^{\infty}i\omega_n B_n
e^{-i\omega_n(t-t_{\rm start})}
e^{-i\omega_n r_*^S}
\int_{-\infty}^{+\infty} e^{i\omega_n r_*}ke^{-b(r_*-r_*^S)^2}dr_*
\right ] \nn \\
&=& -{\rm Re}\left [\sum_{n=0}^{\infty}i\omega_n k B_n\sqrt{\frac{\pi}{b}}
e^{-\frac{\omega_n^2}{4b}}
e^{-i\omega_n(t-t_{\rm start})}
\right ]\,.
\label{gaussfar}
\eeq
If the gaussian is centered very close to the black hole ($r_*^S \ll 0$) we
have
\be
X^{(0)}(t\,,r)\simeq
-{\rm Re}\left [\sum_{n=0}^{\infty}\frac{i(\omega_n-2m\Omega) k B_n}{A_{\rm
out}}e^{im\Omega r_*^S}
\sqrt{\frac{\pi}{b}}
e^{-\frac{(\omega_n-m\Omega)^2}{4b}}
e^{-i\omega_n(t-t_{\rm start})} \right ]\,.
\label{gaussclose}
\ee
Once again, frame dragging effects show up only for gaussians located near the
horizon. The absolute value of each term in the sum is proportional to
$\exp[-{\rm Re}(w)/4b]$, with $w=\omega_n^2$ or $w=(\omega_n-m\Omega)^2$
depending on the location of the initial data. Minimizing with respect to $b$
we see that the maximal excitation corresponds to $b={\rm Re}(w)/2$: we found
a similar result in the vibrating string example [recall the discussion below
Eq.~(\ref{stringexc})].

In the limit of narrow gaussians we expect to recover the results for
localized initial data. Indeed, if we use the $\delta$-function representation
setting $k=(2\sqrt{\pi \epsilon})^{-1}$ and $b=1/(4\epsilon)$:
\be \delta (r_*-r_*^S)=\frac{1}{2\sqrt{\pi
\epsilon}}e^{-\frac{(r_*-r_*^S)^2}{4\epsilon}}\,,
\ee
by taking the limit $\epsilon\to 0$ we recover (\ref{pointsourcefaraway}) and
(\ref{pointsourceclose}).

\subsection{Effective amplitude and initial data}

The present calculation of the QNEFs is only a first step towards the
determination of the QNM content of a waveform. The degree to which a given
QNM is excited depends crucially, through the integral appearing in
(\ref{exccoeffs}), on the initial data $I(\omega\,,r)$.
Can we identify some physical quantity which is only weakly dependent on the
initial data, and therefore useful to discuss the detectability of QNMs?

Andersson and Glampedakis proposed to use a QNM ``effective amplitude''
defined as follows \cite{kostasnils,quickdirty}.  Consider a typical ringdown
waveform $\Psi$. Matched filtering increases the gravitational wave amplitude
by $\sqrt{N}$, $N\propto {\rm Re}(\omega_n)/{\rm Im}(\omega_n)$ being the
number of cycles, yielding an effective amplitude $\Psi_{\rm eff}\sim \sqrt{N}
\Psi$. For a single mode $\Psi \propto C_n$, so that $\Psi_{\rm eff}\sim
\sqrt{N} C_n$.  Now, if we happen to know that the convolution of the initial
data with the homogeneous solution is {\it reasonably independent of
  $\omega_n$}, we could write
\be
\Psi_{\rm eff}\sim \sqrt{{\rm Re}(\omega_n)/{\rm Im}(\omega_n)} B_n \,,
\label{criteriondetect}
\ee
which would be the sought criterion for detectability of a given mode. In
\cite{kostasnils,quickdirty} the real part of (\ref{criteriondetect}) was
referred to as the ``effective gravitational wave amplitude'', and used to
discuss QNM detectability.

The derivation of Eq.~(\ref{criteriondetect}) relies upon the assumption that
the proportionality factor between the $C_n$'s and the $B_n$'s is weakly
dependent on $\omega_n$, for only in this case is the previous expression
valid.  Eqs.~(\ref{pointsourcefaraway}) and (\ref{pointsourceclose}) show that
such an assumption is reasonable for localized initial data.  Unfortunately
this won't be true in more general situations.
A simple counterexample is provided by gaussian initial data. According to
(\ref{criteriondetect}) QNM detectability should be weakly dependent on
$\omega_n$ (at least for low-lying modes), but we know for a fact that the
correct answer for gaussian initial data, Eq.~(\ref{gaussfar}), has an {\it
  exponential} dependence on $\omega_n$. Any notion of detectability based on
the naive argument leading to (\ref{criteriondetect}) is therefore dangerous,
and a case-by-case analysis of the initial data is mandatory.  A study of the
black hole's response to physically reasonable initial data will be the
subject of future work.

\section{Conclusions and outlook}
\label{conclusions}

Motivated by the prospect to test the Kerr nature of astrophysical black holes
by gravitational wave observations, in this paper we compute QNEFs for
general-spin perturbations of Kerr black holes.  For corotating modes with
$l=m$ we find that QNEFs tend to zero in the extremal limit, and that the
overtone contribution is more significant when the black hole is fast
rotating. This result is confirmed by numerical time-evolutions of scalar
perturbations of Kerr black holes \cite{Dorband:2006gg}. In Appendix
\ref{app:asymptBn} we also present the first analytical calculation of the
large-$n$ QNEFs for static black holes, including the Schwarzschild and
Reissner-Nordstr\"om metrics.

QNEFs are ``universal'' properties of the Kerr metric: they do not depend on
the physical nature of the perturbation. Suppose that a distorted Kerr black
hole is formed as a result of a binary merger.  The detailed physics of the
merger, depending on parameters such as the masses, spin magnitude and
inclination of the binary members, will affect the relative overtone
excitation through the initial data function $I(\omega\,,r)$ appearing in the
definition (\ref{exccoeffs}) of the excitation coefficients. Therefore the
problem of determining the relative overtone excitation in ringdown waveforms
reduces to the determination of the function $I(\omega\,,r)$ from numerical
simulations.

Of course the present formalims relies on the possibility to use {\it linear}
perturbation theory to extract ringdown waveforms from numerical simulations.
In the nonlinear regime the total mass and angular momentum of the dynamical
spacetime could deviate significantly from their values for the final Kerr
black hole, introducing systematic (and perhaps time-dependent) redshifts in
the QNM spectrum \cite{Zlochower:2003yh}. An invariant criterion to monitor
the applicability of linear perturbation theory is to use the Petrov
classification, looking at the relative deviation of the background spacetime
from Type D. The Petrov type of the background can be determined, for example,
using a wave-extraction formalism based on the quasi-Kinnersley tetrad and
looking at deviations of the scalar invariant $S$ from the value ($S=1$) that
it would have for a Kerr black hole \cite{Campanelli:2005ia}.

A significant technical challenge in the calculation of the excitation
coefficients $C_n$ is the renormalization of the divergent integral
(\ref{exccoeffs}) for generic initial data. We will address this problem in
future publications. As a testbed we plan to study simple models amenable to
perturbation theory (eg. the old problem of particles plunging into Kerr black
holes). Then we wish to explore the nonlinear regime in relatively symmetric
situations, such as head-on collisions or nonlinear simulations of single,
distorted black holes. Finally we will consider the harder, astrophysically
realistic problem of determining the relative overtone excitation from
nonlinear simulations of merging black holes, using Newman-Penrose based wave
extraction of the ringdown waveform from full numerical relativity
simulations.

An accurate determination of the relative QNM excitation in realistic binary
mergers would have interesting astrophysical applications. For example it
could be used to assess the feasibility of tests of the no-hair theorem with
Earth-based and space-based gravitational wave interferometers, and to improve
present estimates of the ``ringdown braking'' phase observed in numerical and
analytical calculations of gravitational wave recoil
\cite{Baker:2006vn,Damour:2006tr}.

\section*{Acknowledgements}

We thank Amanda Gonzales and Ana Sousa for useful discussions. We are grateful
to Nils Andersson, Kostas Glampedakis, Cliff Will, the relativity group at
Louisiana State University and the participants of the workshop on
Astrophysical Applications of Numerical Relativity in Guanajuato (Mexico) for
their comments and constructive criticism. VC acknowledges financial support
from the Funda\c c\~ao Calouste Gulbenkian through Programa Gulbenkian de
Est\'{\i}mulo \`a Investiga\c c\~ao Cient\'{\i}fica. This work was supported
in part by the National Science Foundation under grant PHY 03-53180, and by
NASA grant NNG06GI60.

\appendix

\section{Asymptotic quasinormal excitation factors of static black holes}
\label{app:asymptBn}

One of the fundamental prerequisites in order to have a well-defined QNM
expansion is that the QNEFs $B_n$ ``converge'' in some sense. So far this
problem has not received much attention.  A numerical study of the scalar
QNEFs for large $n$ and non-rotating black holes can be found in
\cite{nils97}. In this Appendix we study {\it analytically} the large-$n$
behavior of the QNEFs for Schwarzschild black holes and other static
geometries. We explicitly compute the dependence of the QNEFs on $n$
(including subleading effects) in three important cases, including the
Schwarzschild and Reissner-Nordstr\"{o}m spacetimes and the widely used
P\"oschl-Teller approximation of the Regge-Wheeler potential. We also show
that for a general class of static spacetimes the leading behavior of the
QNEFs for large $n$ is $\sim 1/n$, and we conjecture that the same asymptotic
behavior should apply to Kerr black holes. It should be easy to generalize our
technique to compute subleading effects. Finally, as a simple counterexample
to the generality of this behavior, we show that the QNEFs for a potential
barrier scale as $1/n^3$ for large $n$.

\subsection{Schwarzschild black holes}

Our general method relies on the monodromy argument by Motl and Neitzke
\cite{motlneitzke}, devised to compute highly damped (i.e. large-$n$) QNM
frequencies (see \cite{bertiqnm} and \cite{cardosoreview} for reviews). Using
this method Neitzke computed the reflection and transmission coefficients $R$
and $T$ for large $n$ \cite{neitzke}. From his calculation we can simply read
off the in-going and out-going wave amplitudes for the Regge-Wheeler equation:
\beq
A_{\rm in}&=&\frac{1}{T}= \frac{e^{\beta \omega}+1+2\cos(\pi s)}{e^{\beta \omega}-1}\,, \\
A_{\rm out}&=&\frac{R}{T}= \frac{2i\cos{(\pi s/2)}}{e^{\beta \omega}-1}\,, \eeq
where in our units ($G=c=2M=1$) $\beta=4\pi$ is the inverse of the black
hole's Hawking temperature. This result is valid for large $n$ and for {\it
  any} angular quantum number $l$. QNM frequencies, being poles of the
reflection and transmission coefficients, satisfy
\be e^{\beta \omega_{n}}+1+2\cos(\pi s)=0\,. \ee
The calculation of the QNEFs $B_n$ for large $n$ is now straightforward.
Denoting the large-$n$ limit of the QNEFs for spin-$s$ perturbations by
$B^{(s)}_n$ we find
\be B^{(s)}_n\equiv
\left ( \frac{A_{\rm out}}{2\omega\partial_{\omega}A_{\rm in}} \right )_{\omega=\omega_n}=
\frac{i\cos{(\pi s/2)}}{\beta \omega_{n} e^{\beta \omega_{n}}}\,. \ee
This leads to the result anticipated in the main text,
Eq.~(\ref{bnasbehavior}):
\be
B^{(0)}_n=B^{(-2)}_n=
-\frac{i}{3\left [\pm \log3-(2n+1)\pi i\right ]}\,,
\qquad (n\to \infty)\,.
\ee
This is our main result for the Schwarzschild QNEFs in the large damping
limit. The analytic prediction agrees very well with Figure 2 in
\cite{nils97}, where the QNEFs are plotted for the first $\sim 200$ scalar
modes. The calculation of Ref.~\cite{nils97} shows that for large $n$ QNEFs
are independent of $l$ and drop linearly with $n$, in agreement with our
prediction.  Notice also that for electromagnetic perturbations the asymptotic
QNEFs $B^{(-1)}_n$ would be zero: higher-order corrections must be taken into
account to obtain a nontrivial result.

\subsection{Reissner-Nordstr\"{o}m black holes}

The monodromy technique is easily adapted to other black hole backgrounds
\cite{monodromy}, allowing the computation of the asymptotic QNEFs for a wide
class of geometries. Take for instance the Reissner-Nordstr\"{o}m metric,
describing a black hole with mass $M$ and charge $Q$. We follow Motl and
Neitzke \cite{motlneitzke} and fix units such that $M=(k+1)/2$ and $Q^2=k$.
With this choice the outer horizon is located at $r=1$, and we get
\cite{neitzke}
\beq
A_{\rm in}&=&\frac{1}{T}= \frac{e^{\beta \omega}+2+3e^{-\beta_I \omega}}{e^{\beta \omega}-1}\,, \\
A_{\rm out}&=&\frac{R}{T}= \pm i\sqrt{3}\,\frac{1+e^{-\beta_I \omega}}{e^{\beta \omega}-1}\,, \label{AoutRN} \eeq
for any massless integer-spin field (at odds with the Schwarzschild case,
where this symmetry is broken for electromagnetic perturbations). Here
$\beta=4\pi/(1-k)$ and $\beta_I=-k^2\beta$ denote the Hawking temperatures of
the outer and inner horizons, respectively. For large $n$ the QNM frequencies
$\omega_n$ are solutions of
\be\label{RNQNM}
e^{\beta \omega}+2+3e^{-\beta_I \omega}=0\,.
\ee
An elementary calculation similar to the Schwarzschild case yields
\be
B^{(s)}_n=
\frac{\pm i\sqrt{3}}{2\omega_n}\frac{1+e^{-\beta_I\omega_n}}{\beta e^{\beta
\omega_n}-3\beta_Ie^{-\beta_I \omega_n}}\,,
\qquad (n\to \infty) \,. \label{bnasbehaviorrn}
\ee
In (\ref{AoutRN}) and (\ref{bnasbehaviorrn}) the plus sign refers to
electromagnetic-gravitational perturbations, and the minus sign refers to
scalar perturbations. The fact that in the large-$n$ limit the predictions for
charged black holes do not reduce to the Schwarzschild predictions as
$Q\rightarrow 0$ is known, and agrees with numerical results
\cite{cardosoberti}. In general, the transcendental equation (\ref{RNQNM})
must be solved numerically to obtain the QNM frequencies. As in the
Schwarzschild case, one can verify that the QNEFs $B_n\sim 1/n$.

\subsection{The P\"{o}schl-Teller potential}

The P\"{o}schl-Teller potential has been used in many studies of black hole
perturbation theory as a useful approximation of the Regge-Wheeler potential.
The idea is to replace the ``true'' potential barrier by an analytic
expression of the form $V=V^{\rm PT}\equiv V_0 \cosh^{-2}\alpha(x-x_0)$, which
has the advantage that many calculations can be carried out in closed form
\cite{poschl,poschl2}. From the analytic expression of the reflection and
transmission coefficients given in \cite{poschl} we get
\beq A_{\rm
out}&=&\frac{\Gamma(1+i\omega/\alpha)\,\Gamma(-i\omega/\alpha)}{\Gamma(1+\beta)\Gamma(-\beta)}\,,
\\
A_{\rm
in}&=&\frac{\Gamma(1+i\omega/\alpha)\,\Gamma(i\omega/\alpha)}{\Gamma(1+\beta+i\omega/\alpha)
\Gamma(-\beta+i\omega/\alpha)}\,,
\eeq
where $\beta=-1/2+\sqrt{1/4-V_0/\alpha^2}$. As usual, the QNM frequencies are
poles of $A_{\rm in}$. They can be obtained imposing (say)
$\Gamma(-\beta+i\omega/\alpha)^{-1}=0$ (or equivalently
$-\beta+i\omega/\alpha=-n$). Since $B_n\sim (\partial_\omega A_{\rm
  in})^{-1}$, the only nonzero contribution to $B_n$ must come from the term
$\left[\Gamma(1+i\omega/\alpha)\,\Gamma(i\omega/\alpha)
  /\Gamma(1+\beta+i\omega/\alpha)\right] \partial_\omega
(1/\Gamma(-\beta+i\omega/\alpha))$. Introducing the polygamma function
$\psi_0$, such that $\Gamma'(x)=\Gamma(x)\psi_0(x)$, the derivative we need is
\be \frac{d}{d\omega}\left [\frac{1}{\Gamma(-\beta+i\omega/\alpha)}\right
]=-\frac{i\psi_0(-n)}{\alpha\Gamma(-n)}=\frac{(-1)^n i \,n!}{\alpha}\,, \ee
and the final result for the asymptotic QNEFs of the P\"oschl-Teller potential
reads
\be B^{\rm PT}_n=\frac{i\alpha(-1)^{n+1}}{2\omega_n
\Gamma(1+\beta)\Gamma(-\beta)}\,\frac{\Gamma(n-\beta)\Gamma(1+2\beta-n)}{n!\Gamma(\beta-n)} \,.\ee
Once again, asymptotically $B^{\rm PT}_n \sim 1/n$.

\subsection{Generic static black hole spacetimes and other possible extensions}

An investigation of generic spacetimes with the present technique should be
carried out case by case \cite{monodromy}. However, if we are only interested
in the leading-order behavior for large $n$ we can content ourselves with a
simple Born approximation to the scattering amplitude (see for instance
\cite{medvedmartinvisser}) to deduce that $\omega \sim -i n$, so that
generically the QNEFs $B_n \sim 1/n$ for large $n$.

The generalization of the present results to Kerr black holes is more
difficult. Previous numerical calculations of the asymptotic QNM frequencies
\cite{cardosoberti,bertiqnm} show that the imaginary part is still
proportional to $n$, so we expect that (to leading order) the Kerr QNEFs
should still scale as $B^{\rm Kerr}_n\sim 1/n$.

\subsection{A potential barrier}

As a last example we consider the QNEFs for a potential barrier of height
$V_0$ and width $\bar x$, i.e. the wave equation is
$\partial^2_x\Psi+[\omega^2-V(x)]\Psi=0\,,$ with potential
\begin{equation}
V(x)=\left\{ \begin{array}{ll}
             V_0   & \mbox{for $0<x<\bar x$}\\
             0    & \mbox{for $x<0$ and $x>\bar x$}\,.
\end{array}\right.
\label{potential}
\end{equation}
Defining $k=\sqrt{\omega^2-V_0^2}$, the solution in each of the three
different regions is
\beq \Psi&=&e^{-i\omega x}\,,\quad x<0\,,\\
\Psi&=&Ae^{ikx}+Be^{-ikx}\,,\quad 0<x<\bar x\,,\\
\Psi&=&Ce^{i\omega x}+De^{-i\omega x}\,,\quad 0<x<\bar x\,. \eeq
The constants $A,~B,~C,~D$ are obtained by imposing continuity of the field
and its derivative at $x=0$ and $x=\bar x$, so that $A+B=1$ and
$-\omega=k(A-B)/(A+B)$. QNM frequencies are such that $D(\omega)=0$.  This
problem was considered by Chandrasekhar and Detweiler \cite{chandradetweiler},
who showed that for large mode number $n$ the QNM frequencies behave as
$(n\pi-i\omega_i)$, with $\omega_i$ a solution to the transcendental equation
$\sqrt{V_0}=-2\omega_ie^{\omega_i/2}$. We have solved the differential
equation numerically, and our results agree very well with this prediction.
For large overtone $n$ we find the analytical result
\beq C &\sim& \frac{V_0}{4k\omega}e^{-i\bar x (\omega+k)}\left (1-e^{2i\bar x k}\right )\,, \\
D &\sim& -\frac{e^{i\bar x (\omega-k)}}{16k\omega^3}\left
(V_0^2e^{2i\bar x \omega}-16\omega^4\right ) \,.
 \eeq
We also get
\be \frac{d D}{d\omega} \sim -\frac{e^{i\bar x (\omega-k)}}{16k\omega^3}\left
(2i\bar x e^{2i\bar x \omega}V_0^2-64\omega^3
\right ) \,.\ee
At the QNM frequencies $V_0^2e^{2i\bar x \omega}-16\omega^4=0$, and we get the
following QNEFs
\be B^{\rm barrier}_n= \frac{C}{2\omega  \frac{d D}{d\omega}}\sim
\frac{V_0}{32i\bar x \omega^3}\,,\label{excitationfactorbarrier}\ee
where we used the fact that in this regime ${\rm Im}(\omega)\rightarrow
-\infty$ and $e^{-i\bar x (\omega+k)} \sim 0$. The result
(\ref{excitationfactorbarrier}) is in good agreement with numerical results
and shows that the $1/n$ behavior, although very general, is not universal,
for in this case the QNEFs scale as $1/n^3$.

\section{Details on the calculation of the quasinormal excitation factors}
\label{app:tech}

This Appendix provides details on the calculation of QNEFs for general-spin
perturbations of Kerr black holes. Most of the material can be found in the
original papers by Leaver \cite{lePRS,leJMP,lePRD} (see also
\cite{cardosoberti,bertiqnm,BCC}), but we find it convenient to summarize here
the equations which are necessary to implement the computational procedure.

\subsection{Expansion of the angular wavefunction}
\label{app:angular}

If we express the angular wavefunction $S_{lm}$ as
\be
S_{lm}(u)=e^{a\omega u}(1+u)^{\frac{|m-s|}{2}}(1-u)^{\frac{|m+s|}{2}}
\sum_{n=0}^{\infty}a^\theta_n(1+u)^n\,,
\ee
the coefficients $a_n^\theta$ must satisfy the following recurrence relation
\beq
\alpha_0^{\theta}a^\theta_1+\beta_0^{\theta}a^\theta_0&=&0\,,\\
\alpha_n^{\theta}a^\theta_{n+1}+\beta_n^{\theta}a^\theta_n+\gamma_{n}^{\theta}a^\theta_{n-1}&=&0\,,
\eeq
where
\beq
\alpha_n^{\theta}&=&-2(n+1)(n+2k_1+1) \,,\\
\beta_n^{\theta}&=&n(n-1)+2n(k_1+k_2+1-2a\omega)-\nonumber \\
& & \left [2a\omega(2k_1+s+1)-(k_1+k_2)(k_1+k_2+1)\right ]-a^2\omega^2-s(s+1)-A_{lm}\,,\\
\gamma_n^{\theta}&=&2a\omega(n+k_1+k_2+s)\,.
\eeq
Regularity of the solution at the boundaries implies that the sequence of
expansion coefficients must be minimal, and that the separation constant
$A_{lm}$ must be a root of the continued fraction
\be\label{angcf}
0=\beta_0^{\theta}-\frac{\alpha_0^{\theta}\gamma_1^{\theta}}{\beta_1^{\theta}-}
\frac{\alpha_1^{\theta}\gamma_2^{\theta}}{\beta_2^{\theta}-}
\frac{\alpha_2^{\theta}\gamma_3^{\theta}}{\beta_3^{\theta}-}...
\ee
or any of its inversions.

\subsection{Jaff\'e expansion of the radial wavefunction}
\label{app:jaffe}

As discussed below Eq.~(\ref{match}), to compute the Kerr QNEFs we need an
accurate representation of the solutions of the radial equation. A convenient
series solution close to the horizon can be found by methods due to Jaff\'e
(see \cite{lePRS}):
\be
R_{r_+}=e^{i\omega r}
(r-r_-)^{-1-s+i\omega+i\sigma_+}
(r-r_+)^{-s-i\sigma_+}
\sum_{n=0}^{\infty}
a^r_n\left (\frac{r-r_+}{r-r_-} \right )^n\,,\label{Rrmaisdet}
\ee
where the notation is the same as in Sec.~\ref{formalism}. The coefficients
$a_n^r$ are normalized so that $a_0^r=1$, consistently with
Eq.~(\ref{Rrp-norm}). They can be obtained from the recurrence relation
\beq
\alpha_0^{r}a^r_1+\beta_0^{r}a^r_0&=&0\,,\\
\alpha_n^{r}a^r_{n+1}+\beta_n^{r}a^r_n+\gamma_{n}^{r}a^r_{n-1}&=&0\,,
\eeq
where
\beq
\alpha_n^{r}&=&n^2+(c_0+1)n+c_0 \,,\\
\beta_n^{r}&=&-2n^2+(c_1+2)n+c_3\,,\\
\gamma_n^{r}&=&n^2+(c_2-3)n+c_4-c_2+2\,,
\eeq
and
\beq
c_0&=&1-s-i\omega-\frac{2i}{b}\left(\frac{\omega}{2}-am\right)\,,\\
c_1&=&-4+2i\omega(2+b)+\frac{4i}{b}\left(\frac{\omega}{2}-am\right)\,,\\
c_2&=&s+3-3i\omega-\frac{2i}{b}\left(\frac{\omega}{2}-am\right)\,,\\
c_3&=&\omega^2(4+2b-a^2)-2am\omega-s-1+(2+b)i\omega-A_{lm}+\frac{4\omega+2i}{b}(\frac{\omega}{2}-am)\,,\\
c_4&=&s+1-2\omega^2-(2s+3)i\omega-\frac{4\omega+2i}{b}\left(\frac{\omega}{2}-am\right)\,.
\eeq
QNM frequencies satisfy a continued-fraction relation analogous to
(\ref{angcf}), with the superscript $\theta$ replaced by a superscript $r$.
Given the QNM frequencies and the corresponding angular eigenvalues $A_{lm}$
it is a simple matter to compute $R_{r_+}$ for any finite, not very large
value of $r$. Unfortunately, the convergence of the Jaff\'e expansion gets
worse for large values of $r$. This is precisely the region where the
wavefunction can be expressed as a sum of in- and out-going components, a
necessary procedure to evaluate $A_{\rm in}^T$ and $A_{\rm out}^T$ and their
derivatives. For this reason we must resort to the Coulomb wavefunction
representation first introduced by Leaver \cite{leJMP,lePRD}, and summarized
below.

\subsection{Coulomb wavefunction expansion of the radial wavefunction}
\label{app:coulomb}

Let us first introduce a new wavefunction $h$, related to the Teukolsky
wavefunction $R$ by
\be\label{rinfinity}
R=(r-r_-)^{-1+i\sigma_+}\,(r-r_+)^{-s-i\sigma_+}\,h(z) \,.
\ee
where $z\equiv \omega (r-r_-)$.  The wavefunction $h$ satisfies the
generalized spheroidal wave equation
\be
z(z-\omega x_0)\left [\partial^2_z\,h+(1-2\eta/z)h \right ]+C_1\omega \partial_z h +
\left (C_2+C_3\omega /z\right )h=0\,,\ee
with
\beq
x_0&=&b\,,\quad \eta=-is-\omega\,,\\
C_1&=&b(1-s-2i\sigma_+)\,,\\
C_2&=&-A_{lm}-s(s+1)+ibs\omega+2\omega^2-a^2\omega^2+b\omega^2\,,\\
C_3&=&b(-1+s+i\omega)(1+i\omega-2i\sigma_+)\,.
\eeq
We next expand $h$ as
\be\label{couexp}
h(z)=\sum_{L=-\infty}^{\infty}a_L\,{\cal U}_{L+\nu}(\eta,z)\,,
\ee
where ${\cal U}_{L+\nu}(\eta,z)$ is any combination of the special functions
$F_{L+\nu}(\eta\,,z)$ and $G_{L+\nu}(\eta\,,z)$, known as Coulomb
wavefunctions (see \cite{stegun,leJMP} for more details on these functions).
We are interested in two particular combinations of Coulomb wavefunctions:
\beq
{\cal U}^+_{L+\nu}(\eta,z)&\equiv&G_{L+\nu}(\eta\,,z)+iF_{L+\nu}(\eta\,,z)\,,\\
{\cal U}^-_{L+\nu}(\eta,z)&\equiv&G_{L+\nu}(\eta\,,z)-iF_{L+\nu}(\eta\,,z)\,.
\eeq
As $z \rightarrow \infty$ these functions have a simple asymptotic behavior
\cite{stegun}:
\be
{\cal U}^{\pm}_{L+\nu}(\eta,z) \sim \left (1\pm \frac{(i\eta-L)(i\eta+L+1)}{2iz}\right )
e^{\pm i \left (z-\eta\log{2z}-L\pi/2+\sigma_{L} \right )}\,,
\label{asymptymame}
\ee
with
\be
\sigma_{L}=
-\frac{i}{2}\log
\frac{\Gamma{\left (L+\nu+1+i\eta\right)}}
{\Gamma{\left (L+\nu+1-i\eta\right )}}\,.
\label{slnu}
\ee
Other properties of these functions are given in Appendix \ref{app:checks}.
The coefficients $a_L$ in (\ref{couexp}) are determined by
\be
\alpha_L a_{L+1}+\beta_L a_{L}+\gamma_L a_{L-1}=0\,,\label{aldet}
\ee
where
\beq
\alpha_L&=&-\frac{\omega\sqrt{(L+1+\nu)^2+\eta^2}}{(L+1+\nu)(2L+2\nu+3)}\left [(L+\nu+1)(L+\nu+2)x_0-C_1(L+\nu+2)-
C_3 \right ] \,,\\
\beta_L&=&(L+\nu)(L+\nu+1)+C_2+\frac{\omega \eta}{(L+\nu)(L+\nu+1)}\left [(L+\nu)(L+\nu+1)x_0-C_1-
C_3 \right ] \,,\\
\gamma_L&=&-\frac{\omega\sqrt{(L+\nu)^2+\eta^2}}{(L+\nu)(2L+2\nu-1)}\left [(L+\nu)(L+\nu-1)x_0-C_1(L+\nu-1)-
C_3 \right ] \,.
\eeq
It is convenient to define the quantities \cite{mano}
\be\label{RLdef}
R_L\equiv \frac{a_L}{a_{L-1}} \,,\quad L_L\equiv \frac{a_L}{a_{L+1}}\,.
\ee
From these definitions it follows that
\be
R_L=-\frac{\gamma_L}{\beta_L+\alpha_LR_{L+1}}\,,\quad
L_L=-\frac{\alpha_L}{\beta_L+\gamma_LL_{L-1}}\,.
\ee
For a minimal solution of the recurrence relation the (as yet undetermined)
parameter $\nu$ must be a root of \cite{leJMP}
\be \label{nudet}
\beta_0=\frac{\alpha_{-1}\gamma_{0}}{\beta_{-1}-}\frac{\alpha_{-2}\gamma_{-1}}{\beta_{-2}-}
\frac{\alpha_{-3}\gamma_{-2}}{\beta_{-3}-}...+
\frac{\alpha_{0}\gamma_{1}}{\beta_{1}-}\frac{\alpha_{1}\gamma_{2}}{\beta_{2}-}\frac{\alpha_{2}\gamma_{3}}{\beta_{3}-}...\,.
\ee
The parameter $\nu$ can also be determined as a solution of the equation
\cite{mano}
\be \label{nudet2}
R_L\,L_{L-1}=1.
\ee
However, as suggested in \cite{fujitatagoshi}, the most convenient numerical
choice is to solve for
\be \label{nudet3}
\beta_L+\alpha_L R_{L+1}+\gamma_L L_{L-1}=0\,.
\ee
We are interested only in the solutions that map to the correct asymptotic
value $\nu=l$ as $\omega\rightarrow 0$ (corrections of order $\omega^2$ can be
found in \cite{mano}). Roots $\nu$ that are integer multiples of $1/2$ are
usually spurious and must be discarded \cite{leJMP}. For slowly damped modes,
physically meaningful roots are most easily obtained setting $L=0$ or $L=-1$
in (\ref{nudet3}). We verified that different numerical procedures yield
excellent agreement on the resulting values of $\nu$. For reference, we list
some of these values along with the corresponding QNM frequencies (for $a=0$)
in Table \ref{tab-nu}.

The procedure to compute the functions $R_{\infty_+}$ and $R_{\infty_-}$ is
now straightforward, at least in principle. Compute the parameter $\nu$
solving (\ref{nudet}), (\ref{nudet2}) or (\ref{nudet3}).  Then compute the
coefficients $a_L$ for positive and negative $L$ from (\ref{RLdef}), setting
for instance $a_0=1$ (the final results are of course unaffected by this or
other normalization choices). Finally, the general solution follows from
(\ref{rinfinity}) and (\ref{couexp}).  Imposing the normalization condition
(\ref{asymp22}) we get
\be\label{final}
R_{\infty_{\pm}}=
\frac{(r-r_-)^{-1+i\sigma_+}\,(r-r_+)^{-s-i\sigma_+}}{K_{\pm}}
\sum_{L=-\infty}^{\infty} a_L {\cal U}^{\pm}_{L+\nu}\,,
\ee
where from (\ref{asymptymame}) the normalization constants read
\be
K_{\pm}=\sum_{L=-\infty}^{\infty}a_Le^{\pm \left [-i\eta\log{(2\omega)}-i(L+\nu)\pi/2+i\sigma_{L}\right ]}\,,
\ee
The nontrivial step in the calculation of (\ref{final}) is the evaluation of
the Coulomb wavefunctions ${\cal U}^{\pm}_{L+\nu}$ for complex arguments.
Computing these functions is particularly tricky for large values of $|L|$,
which must be included for (\ref{final}) to converge with acceptable accuracy.
Typically a good precision (of the order of five or more significant digits)
is achieved summing terms up to $|L|\simeq 15$.

We verified our results in a number of ways. We first computed the functions
${\cal U}^\pm_{L+\nu}(\eta,z)$ using their relation with the confluent
hypergeometric function $U$, Eq.~(125) in \cite{leJMP}. Then we checked their
values using a similar representation in terms of the confluent hypergeometric
function $M$, that can be derived from the integral representation (111) of
\cite{leJMP}. The reason for switching to $U$ and $M$ is that confluent
hypergeometric functions (unlike Coulomb wavefunctions) are implemented in the
present version of Mathematica. Our Mathematica calculations agree very well
with a Fortran subroutine to compute Coulomb wavefunctions for complex
arguments \cite{thompson}, that was eventually used to obtain the results in
this paper. Further checks on our numerical results are discussed in the
following section.

\subsection{Numerical checks of the solutions: some useful identities}
\label{app:checks}

Here we describe a number of consistency checks we performed on our numerical
solutions. We performed the calculations using both Mathematica and a Fortran
code. Our final results on the QNEFs typically agree to one part in $10^4$ or
better.

The convergence of the series expansion (\ref{Rrmaisdet}) was checked via a
direct high-precision integration of the Teukolsky equation. When evaluated in
the matching region (typically we use $r\sim 5-7$) the two methods agree to
one part in $10^{7}$ or better.

Besides checking the validity of the asymptotic expansion (\ref{asymptymame}),
the following equalities proved useful to determine the accuracy of the
functions ${\cal U}_{\pm}^{L+\nu}$:
\beq
& & \frac{d^2{\cal U}^{\pm}_{L+\nu}}{dz^2}+\left (1-\frac{2\eta}{z}-\frac{(L+\nu)(L+\nu+1)}{z^2} \right ){\cal U}^{\pm}_{L+\nu}=0\, \\
& & {\cal U}^+_{L+\nu-1}{\cal U}^-_{L+\nu}-{\cal U}^+_{L+\nu}{\cal U}^-_{L+\nu-1}=\frac{2i(L+\nu)}{\sqrt{(L+\nu)^2+\eta^2}}\,
\label{sumL}\\
& & (L+\nu)\sqrt{(L+\nu+1)^2+\eta^2}{\cal U}^{\pm}_{L+\nu+1}=\left (2(L+\nu)+1\right )\left
(\eta+\frac{(L+\nu)(L+\nu+1)}{z}\right ){\cal U}^{\pm}_{L+\nu}-\nonumber \\
& & (L+\nu+1)\sqrt{(L+\nu)^2+\eta^2}{\cal U}^{\pm}_{L+\nu-1}\, \\
& & \frac{d{\cal U}^+_{L+\nu}}{dz}{\cal U}^-_{L+\nu}-{\cal U}^+_{L+\nu}\frac{d{\cal U}^-_{L+\nu-1}}{dz}=2i\,\\
& & \left (L+\nu\right )\frac{d{\cal U}^{\pm}_{L+\nu}}{dz}=\sqrt{(L+\nu)^2+\eta^2}{\cal U}^{\pm}_{L+\nu-1}-
\left (\frac{(L+\nu)^2}{z}+\eta\right ){\cal U}^{\pm}_{L+\nu} \, \\
& & (L+\nu+1)\frac{d{\cal U}^{\pm}_{L+\nu}}{dz}=\left (\frac{(L+\nu+1)^2}{z}+\eta \right ){\cal U}^{\pm}_{L+\nu}-
\sqrt{(L+\nu+1)^2+\eta^2}{\cal U}^{\pm}_{L+\nu+1} \,,
\eeq
Numerical calculations of ${\cal U}_{\pm}^{L+\nu}$ usually fail for large
$|L|$. This is true both when we use their representation in terms of
confluent hypergeometrics and when we use the Fortran routines of
Ref.~\cite{thompson}.  In the sum (\ref{final}) we included only those values
of $L$ for which the identity (\ref{sumL}) is satisfied (in modulus) to better
than one part in $10^4$.

An important test on the linear combinations $R_{\infty_\pm}$ is based on the
calculation of their Wronskian. One can prove easily that the Wronskian $W$
between any two solutions of the Teukolsky equation satisfies
\be
W(r)=W(r_0)
\exp\left[
{-\int_{r_0}^r \frac{(s+1)(2x-1)}{(x-r_+)(x-r_-)} dx}
\right]\,,
\ee
where $r_0$ is any fixed point. Considering the two solutions $R_{\infty_+}$
and $R_{\infty_-}$ we can evaluate the constant $W(r_0)$ by computing the
Wronskian at infinity from their asymptotic behavior (\ref{asymp22}), with the
result
\be
W(r_0)=2i\omega\left [ (r_0-r_+)(r_0-r_-)\right ]^{-(s+1)}\,.
\ee
Our wavefunctions satisfy this relation at different selected values of $r_0$.
A final test consists in evaluating $A_{\rm out}$ by a direct integration of
the Teukolsky equation and comparing the results with the method described in
the main text: the agreement is at the level of one percent or better.

\section{Transformation between the Teukolsky and Sasaki-Nakamura
wavefunctions}
\label{app:SN}

In the main text we computed the QNEFs using the Teukolsky formalism. Quite
often it is computationally convenient to use the equivalent formalism
developed for $s=-2$ by Sasaki and Nakamura \cite{sasakinakamura,SN2,tagoshi}
and its generalization for other spins \cite{hughes}. In general, switching
between the two formalisms is not trivial. Fortunately, deriving the
asymptotic behavior of the respective wavefunctions and the relation between
the QNEFs is quite simple, as we show below.

We follow Sasaki and Tagoshi \cite{tagoshi} and denote the SN function by
$X^{(s)}$. At infinity, the asymptotic behavior of $X^{(s)}$ is the same for
all $s$:
\beq
& & \lim_{r \to \infty } X^{(s)} \sim A_{\rm in} e^{-i\omega r_*}+A_{\rm out}e^{i\omega r_*}\,,\label{ps}
\eeq
%
The normalization at the horizon is fixed by our choice for the Teukolsky
function, Eqs.~(\ref{Rrp-norm}) and (\ref{Rrmaisdet}).  We will define the
QNEFs in the SN formalism in the usual way:
\be
B^{(s)}=\left.
\frac{A_{\rm out}}{2\omega}
\left(\frac{d A_{\rm in}}{d\omega}\right)^{-1}
\right|_{\omega=\omega_{n}}\,.
\ee
Below we establish the relation between $B^{(s)}$ and the QNEFs $B^{(s)}_T$ in
the Teukolsky formalism, as computed in the main text.

\subsection{Scalar perturbations}

In the scalar case, the generalized SN function is related to the Teukolsky
radial function $R$ by \cite{hughes}
\begin{equation}\label{TSNs0}
X^{(0)}=(r^2+a^2)^{1/2}R \,,
\end{equation}
and satisfies
\begin{equation}
\frac{d^2}{dr_*^2}\, X^{(0)}-{\cal U}^{(0)}\, X^{(0)}=0\,, \label{mT-eq}
\end{equation}
where
\begin{eqnarray}
{\cal U}^{(0)}=-\frac{K^2-(A_{lm}-2am\omega
+a^2\omega^2)
\Delta}{(r^2+a^2)^2}+G^2+{\frac{d}{dr_*}}G\,,
\end{eqnarray}
with $K=(r^2+a^2)\omega-am$, and $G=r\Delta(r^2+a^2)^{-2}$.

From Eq.~(\ref{TSNs0}) we have immediately
\be
B^{(0)}_{\rm T}=B^{(0)}\,.
\label{convs0}
\ee
The normalization of the Teukolsky function, Eq.~(\ref{Rrp-norm}), implies
that at the horizon
\begin{equation}
\lim_{r \to r_+} X^{(0)}
\sim (r_+^2+a^2)^{1/2} (r_+-r_-)^{-1+i\omega+i\sigma_+}
e^{i\omega r_+}(r-r_+)^{-i\sigma_+}\, .
\label{normsn-s0}
\end{equation}
%

\subsection{Electromagnetic perturbations}

For the electromagnetic case, the generalized SN transformation is
\be X^{(-1)}=\sqrt{\frac{r^2+a^2}{\Delta}}\left (\alpha \,R^{(-1)}+\beta \partial_r \, R^{(-1)}\right ) \,,\ee
where we can choose \cite{hughes}
\be
\alpha=-\frac{r^2+a^2}{r^2}\sqrt{\Delta}\left (\frac{r}{r^2+a^2}+\frac{iK}{\Delta} \right )\,,\qquad
\beta=\frac{r^2+a^2}{r^2}\sqrt{\Delta}\,.
\ee
From the asymptotic behavior of the Teukolsky function
\be \lim_{r \to \infty }\, R^{(-1)} \sim  A_{\rm in}^{\rm T} r^{-1} e^{-i\omega r_*}\left (1+\mu_1/r\right
)+A_{\rm out}^{\rm T} re^{i\omega r_*}\left (1+\mu_2/r\right )\,,\ee
with
$\mu_1=\left[-2\omega+i\left (2-A_{lm}-a^2\omega^2\right )\right]/(2\omega)$
and
$\mu_2=\left[i\left(A_{lm}+a^2\omega^2\right )\right]/(2\omega)$
we get
\be
A_{\rm in}^{\rm T}=-\frac{1}{2i\omega}A_{\rm in}\,,\qquad
A_{\rm out}^{\rm T}=
-\frac{2i\omega}{2am\omega-A_{lm}-a^2\omega^2}A_{\rm out}\,,
\ee
and finally
\be
B^{(-1)}_{\rm T}=-\frac{4\omega^2}{2am\omega-A_{lm}-a^2\omega^2}B^{(-1)}\,.
\label{convs1}
\ee
Similarly, from the asymptotic behavior of the Teukolsky function at the
horizon (\ref{Rrp-norm}) we find
\be
\lim_{r \to r_+} X^{(-1)}=
\left (1-2i\sigma_+\right ) r_+^{-1/2}
(r_+-r_-)^{i\omega+i\sigma_+} e^{i\omega r_+}(r-r_+)^{-i\sigma_+}\,.
\label{normsn-s1}
\ee

\subsection{Gravitational perturbations}

For the gravitational case the transformation between the SN and Teukolsky
wavefunctions is more complex, but here we only need the asymptotic behavior
(for more details we refer to the original papers \cite{sasakinakamura,SN2}
and to the review in \cite{tagoshi}). One can show, eg. from Eqs.~(65)-(69) in
\cite{tagoshi}, that asymptotically the Teukolsky and SN amplitudes satisfy
\be
A_{\rm in}^{\rm T}=-\frac{1}{4\omega^2}A_{\rm in}\,,\qquad
A_{\rm out}^{\rm T}=-\frac{4\omega^2}
{\lambda(\lambda+2)-6i\omega-12a\omega(a\omega-m)}
A_{\rm out}\,,
\ee
so that
\be
B^{(-2)}_{\rm T}=
\frac{16\omega^4}
{\lambda(\lambda+2)-6i\omega-12a\omega(a\omega-m)}
B^{(-2)}\,.
\label{convs2}
\ee
with $\lambda\equiv A_{lm}+(a\omega)^2-2am\omega$.  Expressing Eqs.~(65) and
(68) in \cite{tagoshi} in Leaver's units and combining them with
Eq.~(\ref{Rrp-norm}), the normalization at the horizon is:
\begin{equation}
\lim_{r \to r_+} X^{(-2)}=
d
(r_+-r_-)^{-1+i\omega+i\sigma_+}
e^{i\omega r_+}(r-r_+)^{-i\sigma_+}\,,
\label{normsn-s2}
\end{equation}
where $d\equiv r_+^{1/2} \left[ (8-12i\omega-4\omega^2)r_+^2+
(12iam-8+8am\omega+6i\omega)r_+ -4a^2m^2-6iam+2 \right]$.



\begin{table}[t]
  \centering \caption{\label{tab:omegas0} QNM frequencies of the first
    two overtones with $s=0$ and $l=2$, for several values of rotation
    parameter and for some values of $m$. For consistency with \cite{BCW},
    with Table 3 of \cite{quickdirty} and with most of the QNM literature,
    here we depart from Leaver's unit convention (but not from his Fourier
    transform convention) listing the dimensionless frequencies $\omega
    M$.}
\begin{tabular}{cccccc}
\multicolumn{6}{c}{$s=0$, $n=0$} \\
\hline
$j$ &$m=-2$   &$m=-1$    &$m=0$     & $m=1$   & $m=2$     \\
\hline
%
0.00& 0.48364-0.09676$i$& 0.48364-0.09676$i$& 0.48364-0.09676$i$& 0.48364-0.09676$i$& 0.48364-0.09676$i$\\
0.20& 0.45620-0.09655$i$& 0.47015-0.09650$i$& 0.48491-0.09646$i$& 0.50055-0.09641$i$& 0.51712-0.09638$i$\\
0.40& 0.43306-0.09599$i$& 0.45932-0.09577$i$& 0.48886-0.09547$i$& 0.52214-0.09515$i$& 0.55964-0.09493$i$\\
0.50& 0.42275-0.09562$i$& 0.45477-0.09523$i$& 0.49196-0.09463$i$& 0.53536-0.09397$i$& 0.58599-0.09349$i$\\
0.60& 0.41315-0.09520$i$& 0.45073-0.09457$i$& 0.49594-0.09348$i$& 0.55080-0.09219$i$& 0.61737-0.09125$i$\\
0.80& 0.39573-0.09429$i$& 0.44407-0.09280$i$& 0.50713-0.08967$i$& 0.59202-0.08513$i$& 0.70683-0.08152$i$\\
0.90& 0.38780-0.09379$i$& 0.44141-0.09165$i$& 0.51478-0.08641$i$& 0.62183-0.07718$i$& 0.78164-0.06929$i$\\
0.98& 0.38177-0.09338$i$& 0.43957-0.09057$i$& 0.52212-0.08254$i$& 0.65470-0.06290$i$& 0.89802-0.04090$i$\\
\hline
\hline
\multicolumn{6}{c}{} \\
\multicolumn{6}{c}{$s=0$, $n=1$} \\
\hline
$j$ &$m=-2$   &$m=-1$    &$m=0$     & $m=1$   & $m=2$     \\
\hline
%
0.00& 0.46385-0.29560$i$& 0.46385-0.29560$i$& 0.46385-0.29560$i$& 0.46385-0.29560$i$& 0.46385-0.29560$i$\\
0.20& 0.43418-0.29576$i$& 0.44940-0.29518$i$& 0.46539-0.29460$i$& 0.48221-0.29405$i$& 0.49989-0.29358$i$\\
0.40& 0.40907-0.29477$i$& 0.43805-0.29314$i$& 0.47013-0.29132$i$& 0.50578-0.28955$i$& 0.54541-0.28821$i$\\
0.50& 0.39787-0.29398$i$& 0.43337-0.29153$i$& 0.47381-0.28856$i$& 0.52020-0.28555$i$& 0.57344-0.28334$i$\\
0.60& 0.38740-0.29304$i$& 0.42927-0.28950$i$& 0.47843-0.28476$i$& 0.53697-0.27969$i$& 0.60667-0.27599$i$\\
0.80& 0.36839-0.29081$i$& 0.42253-0.28395$i$& 0.49060-0.27224$i$& 0.58067-0.25688$i$& 0.70042-0.24548$i$\\
0.90& 0.35971-0.28956$i$& 0.41969-0.28029$i$& 0.49759-0.26177$i$& 0.60984-0.23150$i$& 0.77768-0.20801$i$\\
0.98& 0.35310-0.28851$i$& 0.41754-0.27687$i$& 0.50230-0.25024$i$& 0.62982-0.18739$i$& 0.89622-0.12214$i$\\
\hline
\hline
\end{tabular}
\end{table}

\begin{table}[t]
  \centering \caption{\label{tab:Alms0} Angular separation constants
    $A_{lm}$ of the first two overtones with $s=0$ and $l=2$, for several
    values of rotation parameter and for some values of $m$.}
\begin{tabular}{cccccc}
\multicolumn{6}{c}{$s=0$, $n=0$} \\
\hline
$j$ &$m=-2$   &$m=-1$    &$m=0$     & $m=1$   & $m=2$     \\
\hline
0.00& 6.0000+0.00000$i$& 6.0000+0.00000$i$& 6.0000+0.00000$i$& 6.0000+0.00000$i$& 6.0000+0.00000$i$\\
0.20& 5.9989+0.00050$i$& 5.9964+0.00156$i$& 5.9953+0.00196$i$& 5.9959+0.00166$i$& 5.9985+0.00057$i$\\
0.40& 5.9959+0.00190$i$& 5.9862+0.00604$i$& 5.9807+0.00781$i$& 5.9819+0.00682$i$& 5.9930+0.00243$i$\\
0.50& 5.9939+0.00289$i$& 5.9788+0.00929$i$& 5.9695+0.01217$i$& 5.9702+0.01080$i$& 5.9880+0.00392$i$\\
0.60& 5.9917+0.00405$i$& 5.9700+0.01317$i$& 5.9553+0.01743$i$& 5.9544+0.01570$i$& 5.9808+0.00582$i$\\
0.80& 5.9865+0.00684$i$& 5.9482+0.02266$i$& 5.9167+0.03030$i$& 5.9056+0.02776$i$& 5.9547+0.01063$i$\\
0.90& 5.9836+0.00845$i$& 5.9352+0.02817$i$& 5.8911+0.03744$i$& 5.8674+0.03351$i$& 5.9294+0.01271$i$\\
0.98& 5.9812+0.00982$i$& 5.9237+0.03289$i$& 5.8668+0.04293$i$& 5.8245+0.03416$i$& 5.8884+0.01030$i$\\
\hline
\hline
\multicolumn{6}{c}{} \\
\multicolumn{6}{c}{$s=0$, $n=1$} \\
\hline
$j$ &$m=-2$   &$m=-1$    &$m=0$     & $m=1$   & $m=2$     \\
\hline
0.00& 6.0000+0.00000$i$& 6.0000+0.00000$i$& 6.0000+0.00000$i$& 6.0000+0.00000$i$& 6.0000+0.00000$i$\\
0.20& 5.9994+0.00147$i$& 5.9980+0.00455$i$& 5.9973+0.00575$i$& 5.9975+0.00486$i$& 5.9991+0.00168$i$\\
0.40& 5.9982+0.00552$i$& 5.9927+0.01762$i$& 5.9886+0.02294$i$& 5.9882+0.02010$i$& 5.9951+0.00719$i$\\
0.50& 5.9974+0.00836$i$& 5.9890+0.02709$i$& 5.9815+0.03577$i$& 5.9797+0.03186$i$& 5.9911+0.01163$i$\\
0.60& 5.9967+0.01169$i$& 5.9845+0.03838$i$& 5.9720+0.05129$i$& 5.9676+0.04641$i$& 5.9850+0.01727$i$\\
0.80& 5.9954+0.01961$i$& 5.9732+0.06591$i$& 5.9439+0.08920$i$& 5.9256+0.08210$i$& 5.9606+0.03168$i$\\
0.90& 5.9948+0.02414$i$& 5.9662+0.08181$i$& 5.9237+0.10993$i$& 5.8894+0.09850$i$& 5.9347+0.03791$i$\\
0.98& 5.9944+0.02799$i$& 5.9599+0.09535$i$& 5.9043+0.12559$i$& 5.8509+0.09778$i$& 5.8908+0.03067$i$\\
\hline
\hline
\end{tabular}
\end{table}

\begin{table}[t]
  \centering \caption{\label{tab:scalarexccoef} QNEFs $B^{(0)}$ of the first
    two overtones with $s=0$ and $l=2$, for several values of rotation
    parameter and for some values of $m$.
  }
\begin{tabular}{cccccc}
\multicolumn{6}{c}{$s=0$, $n=0$} \\
\hline
$j$ &$m=-2$   &$m=-1$    &$m=0$     & $m=1$   & $m=2$     \\
\hline
0.00& 0.11936+0.01343$i$& 0.11936+0.01343$i$& 0.11936+0.01343$i$& 0.11936+0.01343$i$& 0.11936+0.01343$i$\\
0.20& 0.12028-0.00281$i$& 0.12013+0.00695$i$& 0.11907+0.01746$i$& 0.11687+0.02868$i$& 0.11332+0.04050$i$\\
0.40& 0.12079-0.01047$i$& 0.12090+0.00828$i$& 0.11714+0.03012$i$& 0.10769+0.05442$i$& 0.09045+0.07951$i$\\
0.50& 0.12156-0.01146$i$& 0.12130+0.01197$i$& 0.11447+0.04016$i$& 0.09715+0.07141$i$& 0.06518+0.10076$i$\\
0.60& 0.12283-0.01049$i$& 0.12149+0.01794$i$& 0.10945+0.05296$i$& 0.07919+0.09011$i$& 0.02441+0.11621$i$\\
0.80& 0.12720-0.00092$i$& 0.11939+0.03899$i$& 0.08459+0.08650$i$& 0.00372+0.11379$i$&-0.09556+0.05169$i$\\
0.90& 0.13027+0.01075$i$& 0.11456+0.05709$i$& 0.05691+0.10467$i$&-0.05609+0.08585$i$&-0.05739-0.07009$i$\\
0.98& 0.13253+0.03151$i$& 0.10313+0.08251$i$& 0.01644+0.11542$i$&-0.07590+0.01089$i$& 0.03015+0.03247$i$\\
\hline
\hline
\multicolumn{6}{c}{} \\
\multicolumn{6}{c}{$s=0$, $n=1$} \\
\hline
$j$ &$m=-2$   &$m=-1$    &$m=0$     & $m=1$   & $m=2$     \\
\hline
0.00& 0.03552-0.26427$i$& 0.03552-0.26427$i$& 0.03552-0.26427$i$& 0.03552-0.26427$i$& 0.03552-0.26427$i$\\
0.20& 0.00031-0.23879$i$& 0.02062-0.25361$i$& 0.04536-0.26760$i$& 0.07508-0.27998$i$& 0.11026-0.28969$i$\\
0.40&-0.01305-0.22194$i$& 0.02373-0.25063$i$& 0.07830-0.27535$i$& 0.15482-0.28823$i$& 0.25502-0.27656$i$\\
0.50&-0.01418-0.21735$i$& 0.03186-0.25195$i$& 0.10651-0.27817$i$& 0.21694-0.27760$i$& 0.36021-0.21856$i$\\
0.60&-0.01204-0.21552$i$& 0.04520-0.25482$i$& 0.14519-0.27668$i$& 0.29610-0.24132$i$& 0.46944-0.08037$i$\\
0.80& 0.00433-0.22175$i$& 0.09487-0.26260$i$& 0.26097-0.23285$i$& 0.44182+0.01597$i$& 0.24418+0.54878$i$\\
0.90& 0.02430-0.23258$i$& 0.14156-0.26409$i$& 0.33334-0.16103$i$& 0.33715+0.26654$i$&-0.50193+0.31383$i$\\
0.98& 0.06291-0.25169$i$& 0.21530-0.25651$i$& 0.39147-0.04418$i$& 0.06435+0.29230$i$& 0.23752-0.12729$i$\\
\hline
\hline
\end{tabular}
\end{table}

\begin{table}[t]
  \centering \caption{\label{tab:scalarexccoef2} Amplitudes $A_{\rm
      out}$ of the first two overtones with $s=0$ and $l=2$, for several
    values of rotation parameter and for some values of $(l\,,m)$.}
\begin{tabular}{cccccc}
\multicolumn{6}{c}{$s=0$, $n=0$} \\
\hline
$j$ &$m=-2$   &$m=-1$    &$m=0$     & $m=1$   & $m=2$     \\
\hline
0.00& 1.1472-1.1426$i$& 1.1472-1.1426$i$& 1.1472-1.1426$i$& 1.1472-1.1426$i$& 1.1472-1.1426$i$\\
0.20& 1.1068-1.2514$i$& 1.1252-1.2067$i$& 1.1455-1.1601$i$& 1.1677-1.1115$i$& 1.1915-1.0608$i$\\
0.40& 1.0516-1.4122$i$& 1.0888-1.3208$i$& 1.1353-1.2211$i$& 1.1904-1.1119$i$& 1.2522-0.9921$i$\\
0.50& 1.0057-1.5269$i$& 1.0547-1.4093$i$& 1.1203-1.2776$i$& 1.2013-1.1287$i$& 1.2930-0.9595$i$\\
0.60& 0.9311-1.6802$i$& 0.9955-1.5331$i$& 1.0886-1.3636$i$& 1.2085-1.1641$i$& 1.3444-0.9272$i$\\
0.80& 0.5060-2.2157$i$& 0.6304-1.9952$i$& 0.8437-1.7289$i$& 1.1526-1.3741$i$& 1.5014-0.8727$i$\\
0.90&-0.3226-2.7081$i$&-0.1229-2.4419$i$& 0.2580-2.1461$i$& 0.8860-1.7127$i$& 1.6345-0.9057$i$\\
0.98&-3.8026-2.0011$i$&-3.3378-1.6360$i$&-2.6005-1.6232$i$&-1.1381-2.0428$i$& 1.7179-1.4791$i$\\
\hline
\hline
\multicolumn{6}{c}{} \\
\multicolumn{6}{c}{$s=0$, $n=1$} \\
\hline
$j$ &$m=-2$   &$m=-1$    &$m=0$     & $m=1$   & $m=2$     \\
\hline
0.00&-2.9403+3.1839$i$& -2.9403+3.1839$i$& -2.9403+3.1839$i$& -2.9403+3.1839$i$& -2.9403+3.1839$i$\\
0.20&-2.3498+2.9311$i$& -2.6009+3.0411$i$& -2.8990+3.1573$i$& -3.2543+3.2790$i$& -3.6792+3.4045$i$\\
0.40&-1.8312+2.6818$i$& -2.2132+2.8695$i$& -2.7542+3.0781$i$& -3.5300+3.2999$i$& -4.6593+3.5080$i$\\
0.50&-1.5742+2.5630$i$& -1.9831+2.7773$i$& -2.6206+3.0195$i$& -3.6326+3.2719$i$& -5.2745+3.4588$i$\\
0.60&-1.3002+2.4481$i$& -1.7093+2.6818$i$& -2.4198+2.9493$i$& -3.6819+3.2125$i$& -5.9898+3.2826$i$\\
0.80&-0.5767+2.2022$i$& -0.8741+2.4579$i$& -1.5934+2.7666$i$& -3.3396+2.9990$i$& -7.6627+2.1745$i$\\
0.90& 0.0721+1.9681$i$& -0.0536+2.2109$i$& -0.5925+2.5617$i$& -2.4246+2.8572$i$& -8.3413+0.9505$i$\\
0.98& 1.2103+0.9694$i$&  1.3828+0.9603$i$&  1.4079+1.1980$i$&  0.4182+1.9828$i$& -8.0724+0.3919$i$\\
\hline
\hline
\end{tabular}
\end{table}

\begin{table}[t]
  \centering \caption{\label{tab:gravexccoef} QNEFs $B^{(-2)}$ of the
    first two overtones with $s=-2$ and $l=2$, for several values of the
    rotation parameter and all values of $m$.
  }
\begin{tabular}{cccccc}
\multicolumn{6}{c}{Sasaki-Nakamura, $s=-2$, $n=0$} \\
\hline
$j$ &$m=-2$   &$m=-1$    &$m=0$     & $m=1$   & $m=2$     \\
\hline
0.00& 0.12690+0.02032$i$& 0.12690+0.02032$i$& 0.12690+0.02032$i$& 0.12690+0.02032$i$  &   0.12690+0.02032$i$ \\
0.20& 0.13384+0.00380$i$& 0.13066+0.01398$i$& 0.12644+0.02431$i$& 0.12114+0.03460$i$  &   0.11481+0.04465$i$ \\
0.40& 0.13951-0.00575$i$& 0.13409+0.01509$i$& 0.12399+0.03694$i$& 0.10873+0.05810$i$  & 0.089192+0.076810$i$ \\
0.50& 0.14255-0.00798$i$& 0.13574+0.01864$i$& 0.12088+0.04702$i$& 0.096848+0.073395$i$& 0.066032+0.094125$i$ \\
0.60& 0.14598-0.00838$i$& 0.13716+0.02454$i$& 0.11523+0.05998$i$& 0.077845+0.090076$i$&   0.03074+0.10817$i$ \\
0.80& 0.15473-0.00179$i$& 0.13753+0.04581$i$& 0.08743+0.09424$i$& 0.00070+0.10841$i$  &-0.091113+0.061347$i$ \\
0.90& 0.16055+0.00821$i$& 0.13413+0.06426$i$& 0.05603+0.11183$i$&-0.059511+0.072290$i$&-0.061104-0.089943$i$ \\
0.98& 0.16652+0.02731$i$& 0.12487+0.08994$i$& 0.01200+0.11809$i$&-0.056299-0.016701$i$&-0.042248+0.067331$i$ \\
\hline
\hline
\multicolumn{6}{c}{} \\
\multicolumn{6}{c}{Teukolsky, $s=-2$, $n=0$} \\
\hline
$j$ &$m=-2$   &$m=-1$    &$m=0$     & $m=1$   & $m=2$     \\
\hline
0.00& 0.025587-0.016876$i$& 0.025587-0.016876$i$& 0.025587-0.016876$i$& 0.025587-0.016876$i$& 0.025587-0.016876$i$ \\
0.20& 0.015858-0.015534$i$& 0.020659-0.016165$i$& 0.026663-0.016183$i$& 0.034089-0.015199$i$& 0.043137-0.012654$i$ \\
0.40& 0.010593-0.013107$i$& 0.018158-0.014516$i$& 0.029957-0.013582$i$& 0.047365-0.006739$i$& 0.070428+0.012901$i$ \\
0.50& 0.008973-0.011907$i$& 0.017585-0.013461$i$& 0.032435-0.011025$i$& 0.055380+0.003377$i$& 0.081892+0.046677$i$ \\
0.60& 0.007817-0.010778$i$& 0.017385-0.012229$i$& 0.035338-0.007035$i$& 0.062135+0.021050$i$& 0.07301 +0.10619$i$ \\
0.80& 0.006570-0.008740$i$& 0.017957-0.008866$i$& 0.040011+0.008791$i$& 0.034721+0.091829$i$&-0.24081 +0.15010$i$ \\
0.90& 0.006467-0.007753$i$& 0.018690-0.006243$i$& 0.037266+0.023214$i$&-0.05801 +0.10686$i$&-0.13315 -0.45394$i$ \\
0.98& 0.006929-0.006758$i$& 0.019607-0.002616$i$& 0.025180+0.038227$i$&-0.10800 -0.041437$i$&-0.60610 +0.26215$i$ \\
\hline
\hline
\multicolumn{6}{c}{} \\
\multicolumn{6}{c}{Sasaki-Nakamura, $s=-2$, $n=1$} \\
\hline
$j$ &$m=-2$   &$m=-1$    &$m=0$     & $m=1$   & $m=2$     \\
\hline
0.00&  0.04769-0.22379$i$& 0.04769-0.22379$i$& 0.04769-0.22379$i$& 0.04769-0.22379$i$& 0.04769-0.22379$i$ \\
0.20&  0.01275-0.21222$i$& 0.03286-0.22075$i$& 0.05588-0.22700$i$& 0.08149-0.23024$i$& 0.10888-0.22995$i$ \\
0.40& -0.00572-0.20420$i$& 0.03229-0.22415$i$& 0.08389-0.23496$i$& 0.14621-0.22961$i$& 0.20947-0.20386$i$ \\
0.50& -0.01064-0.20281$i$& 0.03752-0.22859$i$& 0.10852-0.23864$i$& 0.19604-0.21680$i$& 0.27571-0.15875$i$ \\
0.60& -0.01302-0.20359$i$& 0.04716-0.23482$i$& 0.14339-0.23886$i$& 0.25927-0.18219$i$& 0.34491-0.07025$i$ \\
0.80& -0.00904-0.21482$i$& 0.08713-0.25154$i$& 0.25286-0.19967$i$& 0.36945+0.05388$i$& 0.29993+0.40490$i$ \\
0.90&  0.00093-0.22925$i$& 0.12656-0.26018$i$& 0.32055-0.12532$i$& 0.23328+0.28598$i$&-0.55384+0.46277$i$ \\
0.98&  0.02390-0.25495$i$& 0.18872-0.26443$i$& 0.35732-0.00915$i$&-0.08578+0.18065$i$& 0.58139+0.37949$i$ \\
\hline
\hline
\multicolumn{6}{c}{} \\
\multicolumn{6}{c}{Teukolsky, $s=-2$, $n=1$} \\
\hline
$j$ &$m=-2$   &$m=-1$    &$m=0$     & $m=1$   & $m=2$     \\
\hline
0.00&-0.081135+0.067726$i$&-0.081135+0.067726$i$&-0.081135+0.067726$i$&-0.081135+0.067726$i$&-0.081135+0.067726$i$ \\
0.20&-0.042052+0.057934$i$&-0.061317+0.063041$i$&-0.086857+0.065808$i$&-0.12023+0.06413$i$&-0.16306+0.05488$i$ \\
0.40&-0.023703+0.046853$i$&-0.053312+0.056703$i$&-0.10514+0.05697$i$&-0.18937+0.02718$i$&-0.30908-0.07767$i$ \\
0.50&-0.018645+0.042289$i$&-0.052550+0.053249$i$&-0.11959+0.04661$i$&-0.23424-0.02592$i$&-0.36754-0.28206$i$ \\
0.60&-0.015312+0.038467$i$&-0.053615+0.049396$i$&-0.13714+0.02823$i$&-0.27023-0.12799$i$&-0.26075-0.66420$i$ \\
0.80&-0.012449+0.033074$i$&-0.061191+0.038443$i$&-0.16317-0.05596$i$&-0.00912-0.52799$i$&  2.1447-0.0490$i$ \\
0.90&-0.012978+0.031609$i$&-0.068071+0.029129$i$&-0.13659-0.13081$i$& 0.60114-0.32896$i$& -2.2114+3.6692$i$ \\
0.98&-0.015950+0.031612$i$&-0.077430+0.015783$i$&-0.05976-0.18580$i$& 0.14397+0.42835$i$&  3.7992+4.8108$i$ \\
\hline
\hline
\end{tabular}
\end{table}

\begin{table}[t]
  \centering \caption{\label{tab:gravexccoef2} Amplitudes $A_{\rm out}$
    of the first two overtones with $s=-2$ and $l=2$, for several values
    of the rotation parameter and all values of $m$.}
\begin{tabular}{cccccc}
\multicolumn{6}{c}{Sasaki-Nakamura, $s=-2$, $n=0$} \\
\hline
$j$ &$m=-2$   &$m=-1$    &$m=0$     & $m=1$   & $m=2$     \\
\hline
0.00&-4.7608+2.3771$i$& -4.7608+2.3771$i$& -4.7608+2.3771$i$& -4.7608+2.3771$i$ & -4.7608+2.3771$i$  \\
0.20&-6.2957+3.7664$i$& -5.4365+3.0202$i$& -4.6676+2.3909$i$& -3.9841+1.8690$i$ & -3.3819+1.4446$i$  \\
0.40&-7.8741+5.8832$i$& -5.9349+3.8929$i$& -4.3699+2.4414$i$& -3.1373+1.4490$i$ & -2.2068+0.8273$i$  \\
0.50&-8.5905+7.3807$i$& -6.0631+4.4740$i$& -4.1247+2.4901$i$& -2.6907+1.2641$i$ & -1.7053+0.6084$i$  \\
0.60&-9.1362+9.3469$i$& -6.0428+5.2133$i$& -3.7910+2.5653$i$& -2.2271+1.0932$i$ & -1.2657+0.4363$i$  \\
0.80& -8.311+16.035$i$& -4.8599+7.6211$i$& -2.6288+2.8690$i$& -1.2248+0.7843$i$ &-0.58015+0.18762$i$ \\
0.90& -3.399+22.656$i$& -2.0780+9.8297$i$& -1.3270+3.1477$i$&-0.63687+0.62642$i$&-0.32078+0.07721$i$ \\
0.98& 23.657+28.208$i$&  9.6770+9.8318$i$&  2.1270+2.4156$i$& 0.03281+0.34446$i$&-0.09492-0.12174$i$ \\
\hline
\hline
\multicolumn{6}{c}{} \\
\multicolumn{6}{c}{Teukolsky, $s=-2$, $n=0$} \\
\hline
$j$ &$m=-2$   &$m=-1$    &$m=0$     & $m=1$   & $m=2$     \\
\hline
0.00& 0.39793-0.36149$i$& 0.39793-0.36149$i$& 0.39793-0.36149$i$& 0.39793-0.36149$i$& 0.39793-0.36149$i$ \\
0.20& 0.38459-0.43342$i$& 0.38883-0.39777$i$& 0.39300-0.36207$i$& 0.39696-0.32633$i$& 0.40050-0.29056$i$ \\
0.40& 0.35498-0.51546$i$& 0.36539-0.44015$i$& 0.37598-0.36462$i$& 0.38528-0.28888$i$& 0.39152-0.21308$i$ \\
0.50& 0.32915-0.56460$i$& 0.34451-0.46621$i$& 0.36044-0.36748$i$& 0.37380-0.26837$i$& 0.38051-0.16934$i$ \\
0.60& 0.28965-0.62309$i$& 0.31268-0.49797$i$& 0.33704-0.37241$i$& 0.35645-0.24626$i$& 0.36232-0.12071$i$ \\
0.80& 0.10472-0.79207$i$& 0.16769-0.59291$i$& 0.23613-0.39539$i$& 0.28646-0.19715$i$& 0.28191-0.00589$i$ \\
0.90&-0.18616-0.91652$i$&-0.05125-0.65940$i$& 0.09557-0.41518$i$& 0.20434-0.17202$i$& 0.19003+0.05215$i$ \\
0.98& -1.2247-0.7046$i$&-0.77025-0.43494$i$&-0.31955-0.27607$i$& 0.02188-0.13092$i$& 0.04947+0.05337$i$ \\
\hline
\hline
\multicolumn{6}{c}{} \\
\multicolumn{6}{c}{Sasaki-Nakamura, $s=-2$, $n=1$} \\
\hline
$j$ &$m=-2$   &$m=-1$    &$m=0$     & $m=1$   & $m=2$     \\
\hline
0.00& 2.2945 -7.1873$i$&  2.2945-7.1873$i$&  2.2945-7.1873$i$&    2.2945-7.1873$i$&  2.2945-7.1873$i$ \\
0.20& 2.5658 -9.0390$i$&  2.3970-8.0165$i$&  2.2247-7.0665$i$&    2.0424-6.1893$i$&  1.8441-5.3859$i$ \\
0.40& 2.543  -10.849$i$&  2.2599-8.6166$i$&  1.9888-6.6818$i$&    1.6731-5.0445$i$&  1.2672-3.7148$i$ \\
0.50& 2.344  -11.695$i$&  2.0427-8.7930$i$&  1.7817-6.3679$i$&    1.4444-4.4156$i$&  0.9446-2.9579$i$ \\
0.60& 1.933  -12.458$i$&  1.6641-8.8477$i$&  1.4852-5.9476$i$&    1.1812-3.7433$i$&  0.6120-2.2745$i$ \\
0.80&-0.205  -13.360$i$& -3.1529-8.2731$i$&  0.4135-4.5896$i$&    0.5162-2.2182$i$&  0.0351-1.2045$i$ \\
0.90&-3.006  -12.697$i$& -1.9831-7.0189$i$& -0.6196-3.3073$i$&    0.0900-1.2838$i$&-0.04965-0.79665$i$ \\
0.98&-8.7106 -7.3992$i$& -5.0880-2.5636$i$& -1.8011-0.6338$i$& -0.19532-0.23025$i$& 0.07333-0.27876$i$ \\
\hline
\hline
\multicolumn{6}{c}{} \\
\multicolumn{6}{c}{Teukolsky, $s=-2$, $n=1$} \\
\hline
0.00& 0.82322+0.75284$i$& 0.82322+0.75284$i$& 0.82322+0.75284$i$& 0.82322+0.75284$i$& 0.82322+0.75284$i$ \\
0.20& 0.84911+0.70754$i$& 0.82947+0.72453$i$& 0.80991+0.74533$i$& 0.79038+0.76996$i$& 0.77073+0.79838$i$ \\
0.40& 0.85814+0.65292$i$& 0.81337+0.67702$i$& 0.76818+0.71734$i$& 0.72213+0.77414$i$& 0.67346+0.84688$i$ \\
0.50& 0.85843+0.61654$i$& 0.79766+0.64004$i$& 0.73512+0.68991$i$& 0.66983+0.76696$i$& 0.59731+0.86954$i$ \\
0.60& 0.85685+0.56852$i$& 0.77713+0.58786$i$& 0.69265+0.64679$i$& 0.60121+0.74743$i$& 0.49254+0.88500$i$ \\
0.80& 0.84873+0.39392$i$& 0.72054+0.39027$i$& 0.57393+0.46234$i$& 0.39631+0.61795$i$& 0.14457+0.81785$i$ \\
0.90& 0.82922+0.19249$i$& 0.66918+0.17074$i$& 0.48666+0.24733$i$& 0.25229+0.42940$i$&-0.10621+0.60268$i$ \\
0.98& 0.65127-0.27305$i$& 0.44118-0.27437$i$& 0.27060-0.14913$i$& 0.10617+0.07862$i$&-0.17382+0.14948$i$ \\
\hline
\hline
\end{tabular}
\end{table}

\begin{table*}
  \caption{Values of the QNM frequencies and of the angular momentum
    parameter $\nu$ in the Schwarzschild limit.}
\begin{tabular}{cccc}
\multicolumn{4}{c}{$\nu$} \\
$n$ &$s=0$, $l=2$&$s=-1$, $l=1$&$s=-2$, $l=2$\\
\hline
0 & 1.64072+0.24770$i$ &  1.85364+0.20371$i$ & 1.72837+0.23396$i$\\
1 &-1.91021-0.44483$i$ & -1.17974-0.29072$i$ &-2.00310-0.37593$i$\\
\hline
\hline
\multicolumn{4}{c}{$M \omega$} \\
$n$ &$s=0$, $l=2$&$s=-1$, $l=1$&$s=-2$, $l=2$\\
\hline
0 &0.48364-0.09676$i$ &0.24826-0.09249$i$&0.37367-0.08896$i$ \\
1 &0.46385-0.29560$i$ &0.21452-0.29367$i$&0.34671-0.27392$i$ \\
\hline
\hline
\end{tabular}
 \label{tab-nu}
\end{table*}


\begin{thebibliography}{99}
\bibitem{commentsn}

In Eq.~(\ref{waveeq-gen}) of Sec.~\ref{string}, the potential $V(x)$
and the source ${\cal S}$ were introduced to reproduce the structure
of Eq.~(\ref{potbh1}). In general, in the SN formalism the wave
equation contains a first derivative of the wavefunction:
%
\be \frac{d^2X}{dr_*^2}-{\cal F}\frac{dX}{dr_*}-{\cal U}X=S\,,\nonumber  \ee
%
where we also allow for the presence of a source term $S$.  This is no
restriction, since a trivial change of variable of the form
%
\be X=\exp\left[\int \f{{\cal F}}{2} dr_*\right] X_2 \nonumber \ee
%
eliminates the first derivative, yielding
%
\be \frac{d^2X_2}{dr_*^2}+ \left(\f{{\cal F}'}{2}-\f{{\cal F}^2}{4}-\cal U\right) X_2= S\exp\left[-\int
\f{{\cal F}}{2} dr_*\right]\,.\nonumber  \ee
\bibitem{noteqd} The main source of disagreement seems to be a bug in
the computer code used in \cite{quickdirty} to compute scalar QNEFs of
Kerr black holes (N. Andersson and K. Glampedakis, private
communication).
\end{thebibliography}
\end{document}